\documentclass[11pt]{article}
\usepackage{amsmath}
\usepackage{amssymb}
\usepackage{cite}
\usepackage{graphicx,bbm,mathrsfs}
\usepackage{nicefrac}

\usepackage{geometry}       
\newcommand{\be}{\begin{equation}}  
\newcommand{\ee}{\end{equation}}  
\newcommand{\bea}{\begin{eqnarray}}  
\newcommand{\eea}{\end{eqnarray}}

\newcommand\lsim{\mathrel{\rlap{\lower4pt\hbox{\hskip1pt$\sim$}}
    \raise1pt\hbox{$<$}}}
\newcommand\gsim{\mathrel{\rlap{\lower4pt\hbox{\hskip1pt$\sim$}}
    \raise1pt\hbox{$>$}}}

\providecommand{\tabularnewline}{\\}

\newcommand{\captionfonts}{\small}

\newcommand{\approptoinn}[2]{\mathrel{\vcenter{
  \offinterlineskip\halign{\hfil$##$\cr
    #1\propto\cr\noalign{\kern2pt}#1\sim\cr\noalign{\kern-2pt}}}}}

\makeatletter  
\long\def\@makecaption#1#2{%
  \vskip\abovecaptionskip
  \sbox\@tempboxa{{\captionfonts #1: #2}}%
  \ifdim \wd\@tempboxa >\hsize
    {\captionfonts #1: #2\par}
  \else
    \hbox to\hsize{\hfil\box\@tempboxa\hfil}%
  \fi
  \vskip\belowcaptionskip}
\makeatother   

\begin{document}

\vspace*{1.2cm}

\begin{center}

\thispagestyle{empty}

{\Large\bf  Quantified Naturalness from Bayesian Statistics }\\[10mm]

{\large S.~Fichet$^{\,a,\,b}$}\\[5mm]

{\it
$^{a}$~International Institute of Physics, UFRN \\ 
Av. Odilon Gomes de Lima, 1722 - Capim Macio - 59078-400 - Natal-RN, Brazil\\
$^{b}$~Laboratoire de Physique Subatomique et de Cosmologie, UJF Grenoble 1, 
CNRS/IN2P3, 53 Avenue des Martyrs, F-38026 Grenoble, France
}

\vspace*{12mm}

\begin{abstract}
\noindent 


We present a formulation of naturalness made in the framework of Bayesian statistics, which unravels the conceptual problems related to previous approaches. Among other things, the relative interpretation of the measure of naturalness turns out to be unambiguously established by Jeffreys' scale. 
Also, the usual sensitivity formulation (so-called Barbieri-Giudice measure) appears to be embedded in our formulation under an extended form. We derive the general sensitivity formula applicable to an arbitrary number of observables. 
Several consequences and developments  are further discussed. As a final illustration, we work out the map of combined fine-tuning associated to the gauge hierarchy problem and neutralino dark matter in a classic supersymmetric model.

\end{abstract}

\end{center}

\clearpage
%

\section{Introduction}

The notions of naturalness and fine-tuning are a center of interest in some domains of theoretical physics, like the theoretical side of particle physics and cosmology. Loosely speaking, these notions refer to the propensity of a  model to reproduce the experimental observations. When they are employed, their effect is to modify our degree of belief in the model examined. Indeed, intuitively, our degree of belief in a model follows its propensity to fullfil experimental constraints.
For instance, when the parameters of a model should be adjusted very precisely to satisfy a constraint, the model is said to suffer from a lack of naturalness, or to have a fine-tuning problem, and this consideration typically decreases our degree of belief in the model. 

But these considerations, even if they are taken to be intuitive by a certain fraction of people, remain fully subjective and unquantified.
To be more precise and eventually extract some objective information from these intuitive observations, two things are necessary. First, it is necessary to define a consistent measure of  naturalness. Second, it is also necessary to have a rule telling how the measure of naturalness should be mapped to our degree of belief. This second point is important,  because the measure of naturalness would not be usable if the subjectivity was not under control.

Several naturalness issues appear in particle physics, with in particular the gauge hierarchy problem, the strong CP puzzle, the flavour puzzle, as well as in cosmology, with the cosmological constant problem, cosmological coincidence, and the flatness problem. The later being resolved by the inflationary theory. We refer to App.\ref{naturalness_issues} for a short reminder on those different issues.

To our knowledge, it is in the context of the gauge hierarchy problem that a measure of naturalness--i.e. of fine-tuning--was first built. Indeed, supersymmetric (SUSY) models solve the gauge hierarchy problem up to a certain degree, leaving a so-called little hierarchy problem\footnote{The little hierarchy problem comes from the tension remaining between electroweak and TeV scales, and is in fact an issue common to a lot of (all?) models of physics beyond the Standard Model.}.
In the seminal papers  \cite{Ellis:1986yg} and \cite{Barbieri:1987fn}, the amount of fine-tuning is defined as the sensitivity of the electroweak scale (characterized by the Z boson mass) with respect to the model parameters. An ad-hoc formula quantifying the fine-tuning is then derived,   
\be
{\textrm{max}}_{i}\left|\frac{\partial \log m^2_Z}{\partial \log \theta_i}\right|~. \label{BG}
\ee
In this context, the formula gives a measure of the amount of cancellation between the SUSY parameters, which are typically $O(\textrm{TeV})$, necessary to reproduce the Z boson mass, one order of magnitude below.

This sensitivity measure (often called Barbieri-Giudice measure) is largely exploited in the SUSY literature. We refer for example to \cite{Cassel:2010px,Strumia:2011dv,Cabrera:2008tj,Allanach:2006jc} for recent work making use of it. However this formulation has also been criticized, either for its limitations, or at the conceptual level. At the conceptual level, maybe the most straightforward remark is  there is no rule connecting the sensitivity measure to our degree of belief. The interpretation of the numbers provided by Eq.~\eqref{BG} is therefore fully subjective.

Several attempts in the literature have been already made to produce alternative definitions, with in particular the papers \cite{Anderson:1994dz} and \cite{Athron:2007ry}. Among other things, the work \cite{Anderson:1994dz} introduces the key notion of probability distribution of parameters, while \cite{Athron:2007ry} also introduces the important notion of volume of parameter space. Other propositions have also been discussed in \cite{Ciafaloni:1996zh,Chan:1997bi,Barbieri:1998uv,Giusti:1998gz}.
However, even though all of these alternative propositions are well motivated and contain interesting ingredients, it is unfortunately still possible to find conceptual problems and criticisms. Overall, one may find that all those measures of fine-tuning are a bit ad-hoc or lack for a robust framework.

It is with the will of offering a solid framework to the notion of naturalness that we present an approach based on Bayesian statistics. 
Among other things, the link between the naturalness measure and the degree of belief will be established from Jeffreys' scale. Also, our approach turns out to contain the sensitivity measure in a generalized form. Once embedded in this framework, the usual limitations and problems of the sensitivity measure vanish. An attempt in that direction has been done in \cite{Cabrera:2008tj}. It is, to our knowledge, the only paper containing this idea. 

The article is organized as follows. Naturalness problems, the sensitivity formulation and its conceptual flaws are reviewed in  a generic way in Section~\ref{sect:sensitivity}. Section~\ref{sect:modelcomp} is devoted to basics of Bayesian model comparison relevant for our purpose, such that our presentation is self-contained from the point of view of Bayesian statistics.
 We then expose the Bayesian approach to naturalness and its implications in Section~\ref{sect:Bayes_FT}. Section~\ref{sect:SUSY} is finally devoted to some application of the results, focusing mainly on the gauge hierarchy problem in supersymmetric models.

\section{Fine-tuning, puzzles and sensitivity} \label{sect:sensitivity}

In this section, we discuss in a generic way naturalness problems and the sensitivity formulation. The presentation is aimingly transverse, and applicable to any naturalness problem. Along these lines, some of the statements might appear weak or lacking of solid definitions. These inconsistencies will be highlighted in the last paragraph. The critical point of view will be adopted only in this last part. The rest of the section is supporting the sensitivity formulation.

Along the section, we will consider a dimensionless quantity $\delta$ defined in a given model $\mathcal{M}$, with parameters $\theta_i$. We will assume that this $\delta$ is subject to experimental constraints (or any other piece of information exterior to the model). 
 In all generality, one can say that a naturalness problem appears when $\delta$ is constrained to values that it is not expected to take. In particular, it can be different from $O(1)$ while it was expected to be $O(1)$, or at the opposite it can be of $O(1)$ while it was not expected to be $O(1)$. For instance, the gauge hierarchy, cosmological constant and strong CP problems enter in that first category, with $\delta$ being $m^2_Z/M^2_{Pl}$, $\rho_\Lambda/M^4_{Pl}$ and $\theta/2\pi$, respectively. The universe flatness problem and cosmological coincidence enter in the second category, with $\delta$ being $\rho/\rho_c$ and  $\rho_\Lambda/\rho_M$, respectively. However, this splitting into two categories is in fact artificial. Indeed, one has always the freedom to transform one in the other by redefining $\delta\rightarrow 1/(\delta-1)$. Therefore, in this section, whatever the naturalness problem is, we will always choose to define $\delta$ as a number unnaturally smaller than one.

In all generality, $\delta$ is a function of the model parameters, $\delta(\theta_i)$. As we are concerned with the values that $\delta$ can potentially take, the dependence with respect to the parameters is crucial. In the limiting case where $\delta$ does not depend at all on the parameters, it is  completely determined by the model $\mathcal{M}$. In that case, $\mathcal{M}$ is totally predictive, or in other words, totally natural. In the opposite direction, the more $\delta$ depends on the $\theta_i$, the more one has to adjust precisely these parameters to satisfy the experimental constraint, or, in other words, the more $\mathcal{M}$ is fine-tuned.

It is then temptating to define a measure of naturalness by making use of the derivative of $\delta$ with respect to the parameters.  An appropriate quantity has to involve the logarithm of $\delta$ to measure a relative variation, and the logarithm of $\theta_i$ to keep the independence with respect to the choice of units, provided that $\theta_i$ is dimensionful. Therefore this measure has to be based on the quantity $|\partial \log \delta / \partial \log \theta_i|$. Two  such measures have been provided (see \cite{Barbieri:1987fn,Ellis:1986yg,Ciafaloni:1996zh,Casas:2004gh}), in the context of the gauge hierarchy problem and its supersymmetric solution:  
\be
c_a={\textrm{max}}_{i}\left|\frac{\partial \log \delta}{\partial \log \theta_i}\right|~~~~~~\textrm{or}~~~~~~
c_b=\sqrt{\sum_i {\left(\frac{\partial \log \delta}{\partial \log \theta_i}\right)^2}}~.
\ee
Whatever the exact definition is, we will denote this kind of quantity as $c$. With this definition, a fully predictive model has $c=0$, and a model requiring infinite fine-tuning has $c\rightarrow\infty$. 

However, in between these two extreme cases, one can also identify a particular threshold, when $c=1$. This is the particular case where $\delta$ is directly an input parameter of $\mathcal{M}$. The strong CP puzzle and the flavour puzzle in the Standard Model, as well as flatness of the universe and cosmic coincidence in the Standard Cosmological Model are all examples of such a case. Depending on the context and on the opinions, this situation is sometimes considered as being a ``puzzle'', and not a ``problem''. However, this kind of consideration is subjective. It depends ultimately on whether the scientist wishes to find a model more natural than the one with $c=1$, or if he is satisfied with that one. In any case, from the strict point of view of sensitivity, $c=1$ appears well as a limit between predictivity and fine-tuning. 

%

Let us propose two toy examples to illustrate the $c$ measure. To explain the smallness of $\delta$, one often has  to invoke ``special cancellations'' between the $\theta_i$. It is for example the case in the gauge hierarchy problem, where cancellations between $O(M_{Pl}^2)$ quantum contributions needs to occur to obtain $m^2_Z$, or the cosmological constant problem, in which cancellations between  $O(M_{Pl}^4)$ quantum contributions have to occur to make  $\Lambda$ vanish.
 Let us sketch this by $\delta\propto1-\theta$, where $\theta$ is a parameter expected to be $O(1)$. To produce $\delta\ll1$, $\theta$ has to be tuned to be close to one. The $c$ measure is then $|\partial \log \delta / \partial \log \theta|=\theta/\delta$. $c$ is proportional to $\theta$, which is $O(1)$, and to $1/\delta$ which grows with the precision of cancellation required. This quantity is thus well measuring the amount of cancellation necessary to get $\delta\ll1$.
 
 The second toy example is the situation where $\delta\propto e^{-\theta}$, with $\theta$ still an $O(1)$ parameter. This case of ``exponential suppression'' appears for example in the Randall-Sundrum setup to solve the gauge hierarchy problem, in inflationary theories to explain why $\rho/\rho_c-1$ is so small, and also in the dimensional transmutation arising when an asymptotically free theory becomes confining in the infrared. In that case, the $c$ measure $|\partial \log \delta / \partial \log \theta|=\theta$. It does  not depend on $\delta$ but only on the order one parameter. Comparing $c=\theta/\delta$ and $c=\theta$,
 one can see that the ``exponential'' model is more natural by a factor $\delta$ with respect to the ``cancellation'' model.

This way of formulating a measure of naturalness using sensitivity seems well justified, even if rather ad-hoc. However, taking a closer look, one can identify several conceptual flaws, more or less linked together, some of them being already obvious in what we write above. 

Firstly, the notion of ``expectation'' for the value of a quantity, that is used along the section, is not rigourously defined. Even if one tries to express things differently, at some point this notion appears and requires a precise definition. Secondly, the notion of parameter space does not appear in this formulation. It is a bit worrying, as we are concerned with all potential values that $\delta$ could take. These two remarks are particularly suggestive of the Bayesian approach which will be presented in the following sections. But the third, worse issue is the following: there is no rule telling us how to interpret the sensitivity measure in terms of a degree of belief. This holds for the absolute interpretation of $c$, and also at the level of the relative interpretation, when one compares two different values of $c$. For example, we said above that $c$ in the exponential toy model is enhanced by a factor $\delta$ with respect to the cancellation model. But does it really mean that our relative degree of belief between the two models should be given by the value $\delta$? Or maybe $\delta^2$, or $\sqrt{\delta}$? Finally, we can notice the freedom of redefinition of $\delta$. For instance, if one redefines $\delta\rightarrow\delta^{100}$, $c$ is scaled by a factor $100$. Given the absence of rule to interpret $c$, this fact does not constitute a problem in itself. Instead, it can be taken as a constraint of consistency. 
That is, it would be good that the interpretation of $c$ varies consistently with a redefinition of $\delta$, so that the conclusions remain unchanged.


\section{Bayesian model comparison} \label{sect:modelcomp}

The aspects of Bayesian statistics relevant for our purpose are briefly reviewed in this section. For any additional details, we refer the reader to  the comprehensive review \cite{Trotta:2008qt} and references therein, and the textbook \cite{dagostini}.

Within the framework of Bayesian statistics, the notion of probability is defined as 
a measure of the degree of belief about a proposition.
On the other hand, one also knows that whatever the definition of probability $p$ is, the axioms of probability theory entail Bayes' law: 
\be
p(A|B)=p(B|A)\frac{p(A)}{p(B)}~,
\ee
which, with any additional true information $I$, takes the form
\be
p(A|B,I)=p(B|A,I)\frac{p(A|I)}{p(B|I)}~.
\ee
This well known result gets a crucial meaning when applied to probability as a degree of belief. Indeed, replacing
$A$ by any hypothesis $H$, and $B$ by the known information available (called $d$ for ``data''), the previous equality becomes
\be\label{bayes_theorem}
p(H|d,I)=p(d|H,I)\frac{p(H|I)}{p(d|I)}~.
\ee

In Eq.~\eqref{bayes_theorem}, $p(H|I)$ is the probability (i.e. the degree of belief) given to the hypothesis without taking the data into account, which is called prior probability, or just ``prior''. $p(H|d, I)$ is the probability of the hypothesis once the data is taken into account, called posterior probability. 
One thus sees that the Bayes formula, applied  to a piece of information $d$ and a hypothesis $H$, tells how our degree of belief in $H$ should be updated in the light of $d$. This is the remaining term $p(d|H, I)/p(d|I)$ which performs this action. $p(d|H, I)$ is the probability of obtaining the data assuming that the hypothesis is true. But taken as a function of $H$, this quantity is not a probability anymore, and is called a likelihood function,  denoted as $\mathcal{L}(H)$.  It has to be normalized by the constant $p(d|I)$, which is called Bayesian evidence. The Bayesian evidence is the sum over all possible realizations of $H$: 
\be p(d|I)=\sum_H p(d|H,I)p(H|I)~. \label{B_evidence}\ee

Two main applications follow from Eq. \eqref{bayes_theorem}: parameter inference and model comparison. We will be interested in the latter for our purpose. For model comparison, it is the Bayesian evidence Eq.~\eqref{B_evidence} which will play the main role.

Let us consider Eq. \eqref{bayes_theorem}, where hypothesis $H$ is ``model $\mathcal{M}$ is true'' and there is no additional proposition $I$. The equation becomes
 \be
p(\mathcal{M}|d)=p(d|\mathcal{M})\frac{p(\mathcal{M})}{p(d)}~.
\ee
Applying it to two models (which can be the same model with two different priors), $\mathcal{M}_0$ and $\mathcal{M}_1$, and eliminating the unknown constant $p(d)$, one obtains the equation 
 \be\label{bayes_discrimination}
\frac{p(\mathcal{M}_0|d)}{p(\mathcal{M}_1|d)}=\frac{p(d|\mathcal{M}_0)}{p(d|\mathcal{M}_1)} \frac{p(\mathcal{M}_0)}{p(\mathcal{M}_1)}~.
\ee
The quantity $\frac{p(\mathcal{M}_0)}{p(\mathcal{M}_1)}$ is called prior odds, while $\frac{p(\mathcal{M}_0|d)}{p(\mathcal{M}_1|d)}$ is called posterior odds. 
The crucial quantity is the ratio of the Bayesian evidences $\frac{p(d|\mathcal{M}_0)}{p(d|\mathcal{M}_1)}$, denoted as $B_{01}$, and called Bayes factor. 

The Bayes factor tells us how the relative degree of belief between two models is updated given information $d$. A Bayes factor larger [smaller] than 1 will favor $\mathcal{M}_0$ [$\mathcal{M}_1$]. Bayes factors are usually interpreted with respect to Jeffreys' scale \cite{jeffreys1939theory}, given in
Table~\ref{tab:jeffreys}. 
This scale is empirically calibrated, with thresholds at values of the odds of  $3:1$, $12:1$
and $150:1$, representing weak, moderate and strong evidence in favour of $\mathcal{M}_0$, respectively.
It can also be convenient to consider the logarithm $\log B_{01}$.

\begin{table}
\label{tab:jeffreys}

\centering
\begin{tabular}{|c|c|c|c|}
\hline 
$\left|\log B_{01}\right|$ & Odds & Probability & Strength of evidence\tabularnewline
\hline
\hline 
$<1.0$ & $\lesssim3:1$ & $<0.750$ & Inconclusive\tabularnewline
\hline 
$1.0$ & $\sim3:1$ & $0.750$ & Weak evidence\tabularnewline
\hline 
$2.5$ & $\sim12:1$ & $0.923$ & Moderate evidence\tabularnewline
\hline 
$5.0$ & $\sim150:1$ & $0.993$ & Strong evidence\tabularnewline
\hline
\end{tabular}\caption{The empirical Jeffreys scale calibrating the odds between model $\mathcal{M}_0$ and model $\mathcal{M}_1$.}

\end{table}

Note that for a model with continuous parameters, the Bayesian evidence Eq.~\eqref{B_evidence} takes an integral form
\be p(d|I)=\int_{\mathcal{D}} p(d|\theta,\mathcal{M})p(\theta|\mathcal{M})~. \label{B_evidence_int}
\ee
It is then the average of the likelihood function over the parameter space $\mathcal{D}$, weighted by the prior density of the parameters within the model $p(\theta|\mathcal{M})$.

Bayesian model comparison tells us how the odds between two models should be modified by taking into account an external piece of information $d$. 
It formalizes  two competing effects: quality of fit and predictivity. The first one is the usual measure of deviation between data and prediction, given  by the likelihood function. The second one is a principle of economy, i.e. a formalization of Occam's razor. It will  enter in the form of notion of volume in the parameter space. Roughly speaking, provided that the volume of parameter space allowed by the likelihood is smaller than the one allowed by priors (i.e. that data is informative), the Bayes factor will favor the model with the smaller prior volume. Or in other words, it favors the model which is the more predictive with respect to data. This notion of volume is closely related to Fisher information, which, in this context, is a measure of the intrinsic amount of information that the likelihood function and priors contain\cite{lehmann}. For our purposes, we will consider the ``observed'' Fisher information, defined as $I\{f\}(x)=\left|{-\frac{\partial^2 \log f}{\partial x^i x^j }}\right|$. For example, Fisher information of a normal density with variance $\sigma^2$ is $1/\sigma^2$, and Fisher information of a uniform density over the volume $V$ is $1/V^2$. In the present work, it will mainly be this second aspect of predictivity that will matter.

To end this section, let us discuss about the prior density $p(\theta|\mathcal{M})$. The choices of both the functional form and the range of the prior density are critical. The range, conservatively, should be taken as wide as possible. It can be crucial to have ranges which are intrinsically bounded, such that prior volumes remain finite. On the other hand, the functional form of the density is often chosen to be the less informative as possible, i.e. the more objective. Several approaches, based on Fisher information (Jeffreys prior) or the Kullback-Leibler divergence (reference priors, see e.g. \cite{berger_ref_prior}) have been elaborated to construct such priors.

In this work, we will make use of the principle of indifference, which is an approach to minimize the amount of subjective information about a problem.  This principle states that our a priori degree of belief about a problem should be invariant under transformations considered as irrelevant for the problem\cite{press,jaynes}.  Applied to continuous variables, this condition constrains the objective densities and can sometimes fully determine them. For example, a change in coordinates $x'=x+a$ should not influence our a prior degree of belief on $x$. This transformation is thus considered as irrelevant. This imposes the condition $p(x+a)=p(x)$, which constrains $p$ to be the uniform density. Another important example is the one of a dimensionful quantity, $\mu$.  The principle of indifference states that our a priori degree of belief $p(\mu)$ should not depend on the choice of units, such that $\mu'=a\mu$ has the same prior as $\mu$. This translates as the condition $p(\mu)= a p(a\mu)$, which sets $p(\mu)\propto \mu^{-1}$, called logarithmic prior since $\mu^{-1}d\mu=d\log\mu$. As a lot of our observables and parameters are dimensionful, this  logarithmic prior will be omnipresent.

\section{Naturalness in a Bayesian framework } \label{sect:Bayes_FT}

We present in this section the Bayesian approach to the notion of naturalness. First, let us set up the notations.  We consider a model $\mathcal{M}$, with a set of dimensionful parameters $\theta=(\theta_1,\ldots,\theta_n)$, spanning the parameter space $\mathcal{D}$ of dimension $n$. We consider a set of $m$ dimensionful observables $\mathcal{O}(\theta)=(\mathcal{O}_1,\ldots,\mathcal{O}_m)$ (with $m\leq n$) predicted by this model, taking measured value $\mathcal{O}_{ex}$ on the subset of the parameter space $\mathcal{D}_{ex}$ of dimension $n-m$. 
Data other than the $\mathcal{O}$ measurement are collectively called $d$, and the likelihood function $p(\mathcal{O}=\mathcal{O}_{ex}|\theta,\mathcal{M})$ is denoted as $\mathcal{L}_{\mathcal{O}}(\theta)$. An amount of precision $\Sigma$ is associated to the measurement of $\mathcal{O}$. It can be for instance the covariance matrix of a multivariate normal law, or it is more generally given by the Fisher information of the likelihood $\mathcal{L}_\mathcal{O}$ (as a function of $\mathcal{O}$), $I\{\mathcal{L_O}\}(\mathcal{O})=\Sigma^{-1}$. 

Calling $\mathcal{O}$ as an ``observable'' is somewhat misleading. In fact, it just has to be a quantity constrained by experimental data (or any other exterior piece of information). Note that compared to the $\delta$ defined in Section \ref{sect:sensitivity}, we let the $\mathcal{O}$ be dimensionful.
 We emphasize that we restrict ourselves to dimensionful observables and parameters for the sake of simplicity. The consequence of this choice is to make appear logarithmic priors everywhere. However, the whole approach is general to any prior. The generalization does not present difficulty, and will be explained in the last subsection. Finally, the reason of the restriction $m\leq n$ is that $m > n$ is similar to $m=n$ from the naturalness point of view. This point will be discussed afterwards, in the last subsection.


\subsection{Probability formulation}

Loosely speaking, naturalness is the propensity of a given model to reproduce the experimental observation. Using the notations we adopted, the usual translation of this idea is
\begin{itemize}
\item \textit{``Sensitivity of $\mathcal{O}$ with respect to $\theta$, in the
vicinity of a point $\theta_{ex}$ belonging to $\mathcal{D}_{ex}$.'' }
\end{itemize}
This leads to the $c$ measure already presented above,
\be
c_{a}={\textrm{max}}_{i}\left|\frac{\partial \log \mathcal{O}}{\partial \log \theta_i}\right|_{\theta=\theta_{ex}}\label{c_deriv1}
\ee
or
\be
c_{b}=\left.\sqrt{\sum_i {\left(\frac{\partial \log \mathcal{O}}{\partial \log \theta_i}\right)^2}}\right|_{\theta=\theta_{ex}}~.\label{c_deriv2}
\ee
However, an alternative formulation for naturalness, arguably as intuitive as the first one, is
\begin{itemize}
\item \textit{``Probability of having $\mathcal{O}=\mathcal{O}_{ex}$ in the
model.''}
\end{itemize}
This is this second formulation which will be our starting point. As it involves a notion of probability, it necessarily has a Bayesian character. 
This formulation is translated as the probability 
\be
p(\mathcal{O}=\mathcal{O}_{ex}|\mathcal{M},d). \label{pOM}
\ee
But we are interested in this quantity as a function of the hypothesis $(\mathcal{M},d)$. Taken as such, it is not a probability, but instead a Bayesian evidence, as defined in Eq.~\eqref{B_evidence}. 
%
Due to the absence of normalization, this evidence alone is not usable. Instead, it has to appear inside a Bayes factor.
As a measure of naturalness, we therefore have to consider a Bayes factor which compares our hypothesis $(\mathcal{M},d)$ to another hypothesis $(\mathcal{M}',d')$,
\begin{equation} \label{BF_def}
B=\frac{p(\mathcal{O}=\mathcal{O}_{ex}|\mathcal{M},d)}{p(\mathcal{O}=\mathcal{O}_{ex}|\mathcal{M}',d')}~.
\end{equation}
This well-defined quantity plays the main role in our approach.


Two comments are in order. First, it is clear that such a measure of naturalness has a relative character. In this framework, comparing the naturalness of two models $\mathcal{M}$ and $\mathcal{M}'$ is certainly possible, but an absolute statement about the naturalness of $\mathcal{M}$ has to be done with care. To do so, $\mathcal{M}'$ would have to be defined such that it constitutes an absolute reference. How this can be realized will be discuss further in the section.
Second, we emphasize that the distinction between the model $\mathcal{M}$ and the data $d$ is artificial. Indeed, $d$ could be as well considered as a part  of $\mathcal{M}$. It just depends on how $\mathcal{M}$ is defined.  It is convenient to keep this separation explicit for the discussion, and to stress that $d$ and $d'$ need not be identical.

Let us now specify what are the different options available for $(\mathcal{M}',d')$. If one takes the two pieces of data to be identical $d=d'$, and apply $B$ to two realistic models $\mathcal{M}$ and $\mathcal{M}'$, it provides a measure of the relative naturalness between these models. In particular, it makes sense to apply $B$ to the same model with two different prior densities. For instance, one can compare the naturalness of two different regions of the parameter space $\mathcal{D}_{ex}$. One can even choose punctual priors, that is priors that select a single point belonging to $\mathcal{D}_{ex}$. In that case, $B$ measures  the relative naturalness between two points of $\mathcal{D}_{ex}$. This brings us back to a local measure of naturalness, just like the $c$ measure. Note that the selection of a single point of the parameter space also happens if pieces of data $d^{(')}$ are sufficiently constraining. A necessary condition for that is to have at least as many observables  in $d^{(')}$ as parameters in $\mathcal{M}^{(')}$.

Now, let the two models be identical, $\mathcal{M}=\mathcal{M}'$, and let the pieces of data be different, $d\neq d'$. $B$ indicates this time what is the change in naturalness induced by going from data $d$ to data $d'$. Following the literature, this kind of quantity may be dubbed as ``naturalness price'' or ``fine-tuning price'' associated to the change of data. A recent work along this line is \cite{Balazs:2012qc}.

Finally, how can $\mathcal{M}'$ be defined such that it constitutes an absolute naturalness reference? In Section \ref{sect:sensitivity}, we already identified the two limiting cases of total predictivity and infinite fine-tuning. We also identified a threshold in between, when observables $\mathcal{O}_{1\ldots m}$ are input parameters. To define an absolute reference for naturalness, this threshold as well as the limit of total predictivity may be employed. This suggests two ways of defining a reference model.  Either one can consider that $\mathcal{M}'$ is an ideal, fully natural model  satisfying $\mathcal{O}=\mathcal{O}_{ex}$ everywhere in its parameter space. We will denote this ideal model as $\mathcal{X}$. Or, $\mathcal{M}'$ may be a hypothetical ``puzzle''model, in which the $\mathcal{O}_{1\ldots m}$ would be directly input parameters. This model will be denoted $\mathcal{P}$. We call it a ``puzzle'' model, because from the point of view of sensitivity, it stands at the threshold between predictivity and fine-tuning.

These different possibilities for $\mathcal{M}'$ and their implications will be discussed later in the section. 
What we obtain up to now is a well-defined measure of naturalness, under the form of a Bayes factor. Unlike in other approaches, a mapping (Jeffreys' scale) between this measure and our degree of belief exists. The measure is therefore usable, and different applications are possible depending on what one defines as being $(\mathcal{M}',d')$. Starting from now, we will continue the development to show that this probability formulation, instead of being  an alternative to the sensitivity formulation, actually embeds it.

\subsection{Apparition of a sensitivity measure}

From this point until the end of the section, it is assumed that the measurement of $\mathcal{O}$ is sufficiently precise, such that one can consider the Laplace approximation of the likelihood function. That is, the log-likelihood can be expanded around a maximum $\theta_{max}\in \mathcal{D}_{ex}$ as
\be
 \left.\log\mathcal{L_O}(\theta)\simeq \log\mathcal{L}_{max} + \frac{\partial^2\log\mathcal{L_O}}{\partial \theta^i \partial \theta^j}\right|_{\theta_{max}}\frac{(\theta^i-\theta^i_{max})(\theta^j-\theta^j_{max})}{2!}~,
\ee
in which the first order derivatives of $\mathcal{L_O}$ vanishes since $\theta_{max}$ is an extremum. This expansion corresponds to approximating the likelihood as a  (multivariate) normal law. Let us re-write the right-handed term by introducing the Jacobian matrix of the observables $J_{\mathcal{O}\,{ij}}=\left(\frac{\partial O_i}{\partial \theta_j} \right)$, 
\be
\left. \frac{\partial^2\log\mathcal{L_O}}{\partial \mathcal{O}^i \partial \mathcal{O}^j} J_{\mathcal{O}\,ik} J_{\mathcal{O}\,jl} \right|_{\theta_{max}} \frac{(\theta^k-\theta^k_{max})(\theta^l-\theta^l_{max})}{2!}~.\label{likelihood1}
\ee
One recognizes in that expression the quantity $\partial^2\log\mathcal{L_O} / \partial \mathcal{O}^i \partial \mathcal{O}^j$, which up to a minus sign is the observed Fisher information associated to the $\mathcal{O}$ measurement, $I\{\mathcal{L_O}\}(\mathcal{O})$.  
We rescale $\mathcal{O}$ by $\mathcal{O}_{ex}$ to make appear a dimensionless Jacobian and a dimensionless Fisher information associated to $\mathcal{O}/\mathcal{O}_{ex}$, such that Eq.~\eqref{likelihood1} becomes
\be
\left. \frac{\partial^2\log\mathcal{L_O}}{\partial \mathcal{\log O}^i \partial \mathcal{\log O}^j} J_{\mathcal{\log O}\,ik} J_{\mathcal{\log O}\,jl} \right|_{\theta_{max}} \frac{(\theta^k-\theta^k_{max})(\theta^l-\theta^l_{max})}{2!}~.
\ee
The dimensionless Fisher information $-\partial^2\log\mathcal{L_O} / \partial \mathcal{\log O}^i \partial \mathcal{\log O}^j$ describes the amount of relative uncertainty associated to $\mathcal{O}$. We will denote it as $\Sigma^{-1}$ from now on.

Given this expansion of $\mathcal{L_O}$, we can reconsider our central quantity, the Bayesian evidence $p(\mathcal{O}=\mathcal{O}_{ex}|\mathcal{M},d)$. 
This evidence can be written as a continuous sum over all the values of the parameters: 
\be p(\mathcal{O}=\mathcal{O}_{ex}|\mathcal{M},d)=\int_\mathcal{D} \mathcal{L_O}(\theta)\, p(\theta)d\theta~.\label{BE_int}\ee
 It is the average of $\mathcal{L_O}(\theta)$ weighted by the prior density of the model parameters $p(\theta)=p(\theta|\mathcal{M},d)$.  
 We will denote the Fisher information associated to this prior density $I\{p(\theta)\}=|V|^{-1}$, and designate $|V|^{1/2}$ as the ``prior volume''. 

Provided that the likelihood is informative with respect to the prior, Eq.~\eqref{BE_int} takes  the form 
\be \label{evidence_laplace}
p(\mathcal{O}=\mathcal{O}_{ex}|\mathcal{M},d)=\mathcal{L}_{max} \frac{\left|\Sigma\right|^{1/2}}{ \left|V\right|^{1/2}}\int_{\mathcal{D}_{ex}}\frac{1}{C}d\sigma(\theta)~.
\ee
Here, $d\sigma(\theta)$ is the induced integration measure on the manifold $\mathcal{D}_{ex}$, and $C$ is the Jacobian factor, 
\be C= \left|\det\left(J_\mathcal{\log O} J_\mathcal{\log O}^t\right)\right|^{1/2} \label{C_def}~.\ee 
From the point of view of Fisher information,  $C$ measures how much information about the parameters $\theta$ is contained in $\mathcal{O}/\mathcal{O}_{ex}$ regardless of the uncertainty $\Sigma$. The interesting fact is that $C$ is a generalized version of the sensitivity measure $c$, such that Eq.~\eqref{evidence_laplace} makes the link between the two formulations of naturalness.

Some remarks are in order. Firstly, Eq.~\eqref{evidence_laplace} holds in the limit where $C|V|^{1/2}\gg |\Sigma|^{1/2}$ over $\mathcal{D}_{ex}$. 
We will designate the likelihood  as informative when this condition is fulfilled.
 When the condition is not satisfied, the overlap between the likelihood and the prior has to be taken into account properly, and the Bayesian evidence tends to $\mathcal{L}_{max}$. Secondly, we emphasize that $J_\mathcal{\log O} J_\mathcal{\log O}^t$ is indeed a $m\times m$ matrix. Its size does not depend on the number of parameters, but on the number of observables. Thirdly, by choosing a punctual prior, or if the other data $d$ are sufficiently constraining, $\mathcal{D}_{ex}$ is reduced to a single point $\theta_0$ and  the integral $\int_{\mathcal{D}_{ex}}C^{-1}d\sigma(\theta)$ is reduced to $\left.C^{-1}\right|_{\theta_0}$. In such a situation, we get closer from the sensitivity definition, which is a local measure.

We also emphasize that the derivatives which appear within $C$ depend on the choice of prior. Indeed, in all generality, these derivatives are performed with respect to the ``prior repartition function'' $G(\theta)$, defined such that $p(\theta)d\theta=d G(\theta)$.
Let us illustrate this fact by considering a single observable and a flat prior on all parameters, restricted to the volume $[a_1,b_1]\times\ldots\times[a_n,b_n]$. The prior volume is $|V|^{1/2}=(b_1-a_1)\times\ldots\times(b_n-a_n)$, and the Jacobian factor is  $C=\sqrt{\sum_i \left|\partial\mathcal{\log O}/\partial \theta_i\right|^2}$. Instead, if one chooses a logarithmic prior for all parameters $p(\theta_i)\propto\theta_i^{-1}$, the prior volume becomes 
$|V|^{1/2}=(\log b_1-\log a_1)\times\ldots\times(\log b_n-\log a_n)$, and the Jacobian factor becomes  $C=\sqrt{\sum_i \left|\partial\mathcal{\log O}/\partial \log\theta_i\right|^2}$. The derivatives in $C$ are then made with respect to $\log\theta$ instead of $\theta$. In the case of dimensionful parameters, the choice of the logarithmic prior has a particular meaning, because it is the more objective prior.


Through these several remarks, we can finally state that $\int_{\mathcal{D}_{ex}}C^{-1}d\sigma(\theta)$  reduces to the $c$ measure of the sensitivity formulation, provided that one considers a single observable,  a single point in $\mathcal{D}_{ex}$, and that one gives logarithmic priors to the parameters.  It is more precisely the expression $c_b$, used in \cite{Casas:2004gh}, which is exactly reproduced. The measure $c_a$ is an  approximation of $c_b$ when one of the component of the gradient dominates. 

Interestingly, the average $\int_{\mathcal{D}_{ex}}C^{-1}d\sigma(\theta)$ has been proposed in \cite{Anderson:1994dz}, in an attempt to normalize the $c$ measure. In our approach this quantity arises naturally, and we also see that in itself it does not help to interpret the $c$ measure. On the other hand, the use of the volume of parameter space has been proposed in \cite{Athron:2007ry}, in an attempt to build an alternative measure.  These different ideas, somewhat intuitive as such, become rigorously usable once they appear together through Eq.~\eqref{evidence_laplace}. 
Our approach also justifies Bayesian studies which introduce $C^{-1}$ as a ``naturalness prior''. This is not new, it was already explained in \cite{Cabrera:2008tj}. However, we add that the prior of the parameters has to correspond to the derivatives made in $C$, to keep the approach consistent.

The factor $C$ is a generalization of the $c$ measure. Among other things, it tells us what is the information content of several, possibly correlated observables. Let us consider the case of two observables. In this case, $C$ is nothing but the norm of the wedge product of the gradients,
\be
C=\left\|\nabla \mathcal{\log O}_1 \wedge \nabla\mathcal{\log O}_2 \right\|\,,
\ee
which is also 
\be
C=\left(\left\|\nabla \mathcal{\log O}_1 \right\|^2 \left\|\nabla \mathcal{\log O}_2 \right\|^2 - \left( \nabla \mathcal{\log O}_1 . \nabla \mathcal{\log O}_2 \right)^2  \right)^{1/2}~.\label{BF_nm}
\ee
 It is instructive to discuss the behaviour of $C$ depending on the correlation between the two observables which is induced by the model. One can rewrite $C$ as 
\be
C=C_1 C_2 \sqrt{1-\rho^2}\,,
\ee
where $C_1$ and $C_2$ are the one-dimensional sensitivities and
 $\rho$ is the correlation in the model, defined  by
\be
\rho=\frac{\left|\nabla \mathcal{\log O}_1 . \nabla \mathcal{\log O}_2 \right|}{\left\|\nabla \mathcal{\log O}_1\right\|\left\|\nabla \mathcal{\log O}_2\right\|}\,.
\ee

 If the observables are independently predicted, the two gradients are orthogonal and thus $\rho=0$. Eq.~\ref{BF_nm} reduces in that case to the product of the one-dimensional $C$ measures. On the contrary, if the observables are correlated within the model, one has $\rho>0$ and $C$ decreases. In the Bayesian point of view, this should be interpreted as the fact that it is more economical for a model to predict correlated observables than independent observables. One may be worried that Eq.~\eqref{BF_nm} tends to zero in the limit of total correlation, when the two observables are linearly dependent. However, the formula does not apply in that limit. Indeed, recall that the condition for having informative data is $C|V|^{1/2}\gg |\Sigma|^{1/2}$. It translates here as an upper bound on the correlation $\rho$,
\be
\rho \ll \sqrt{1-\frac{|\Sigma|}{|V|(C_1 C_2)^2}}\,. \label{condition_2d}
\ee 
  For instance, for a pair of Gaussian measurements, one has  $|\Sigma|^{1/2}= \sigma_1 \sigma_2 \sqrt{1- \rho_{exp}^2}$. When Eq.~\eqref{condition_2d} is not satisfied, the correlation is too large, such that the two observables are not separately informative. That is, instead of two constraints, the model effectively feels a single constraint $\hat{\mathcal{O}}\equiv\mathcal{O}_1$ $\sim\propto \mathcal{O}_2$.  Instead of  Eq.~\eqref{BF_nm}, it is then a one dimensional sensitivity associated to $\hat{\mathcal{O}}$ which appears in the Bayesian evidence Eq.~\eqref{evidence_laplace}.
  This discussion illustrates also the fact that $C$ taken outside of the formula Eq.~\eqref{evidence_laplace} can induce misunderstandings, and has to be interpreted with care.

At this point, puzzling observations can be made about priors and the meaning of logarithms which appear everywhere. 
What is after all the reason for having $\log\mathcal{O}$ in $C$? 
Is it for the sake of making $C$ dimensionless? Or is it for the sake of measuring a relative variation, as we naively stated in Section \ref{sect:sensitivity}?  Here we assumed that $\mathcal{O}$ is dimensionful. By doing so, we avoid this discussion in a first time, since rescaling $\mathcal{O}$ by $\mathcal{O}_{ex}$ makes the $C$ both dimensionless and measuring a relative variation. 
Also, although the $\log$ in $\partial \log \mathcal{O}$ is suggestive of an objective prior, this remains just a way of writing $\partial\mathcal{O}/\mathcal{O}_{ex}$. And, anyway, speaking about a prior for $\mathcal{O}$  actually does not makes sense for the moment, as $\mathcal{O}$ is determined by the parameters. These observations will be resolved when examining the  Bayes factor $B_{\mathcal{MP}}$ in the next subsection.


To summarize, we find that the sensitivity formulation of naturalness turns out to be embedded in the probability formulation. The $c$ measure, Eq.~\eqref{c_deriv2}, turns out to be a particular case of the factor $C$, 
 arising in the Bayesian evidence Eq.~\eqref{evidence_laplace}. The only assumption made to obtain this result is the Laplace approximation, i.e. that the likelihood function can be reduced to a normal law. We will now examine the different Bayes factors $B$
 that can be constructed, and what is the role taken by the $C$ measure.
 


\subsection{The different versions of $B$}

\subsubsection{$B_{\mathcal{MX}}$}

As a warm-up, let us examine the Bayes factor comparing a model $\mathcal{M}$ to the fully natural model $\mathcal{X}$,
\be
B_{\mathcal{MX}}=\frac{p(\mathcal{O}=\mathcal{O}_{ex}|\mathcal{M})}{p(\mathcal{O}=\mathcal{O}_{ex}|\mathcal{X})}~.
\ee
 By definition,  $\mathcal{X}$ satisfies $\mathcal{O}=\mathcal{O}_{ex}$  in all its parameter space. The evidence of this ideal model is thus $p(\mathcal{O}=\mathcal{O}_{ex}|\mathcal{X})=\mathcal{L}_{max}$. Recall that $\mathcal{L}_{max}$ is an overall normalization constant, which will be canceled once we consider the ratio of evidences. Assuming that the data $\mathcal{O}=\mathcal{O}_{ex}$ is informative for $\mathcal{M}$, $B_{\mathcal{MX}}$ takes the form  \be B_{\mathcal{MX}}=|\Sigma|^{1/2}/|V|^{1/2}\int_{\mathcal{D}_{ex}} C^{-1} d\sigma(\theta)~. \label{BMX} \ee

Clearly, since $\mathcal{M}$ cannot be more natural than $\mathcal{X}$, $B_\mathcal{MX}$ cannot be larger than one.  At most, it can tend to one, if $\mathcal{M}$ tends to be an ideal model like $\mathcal{X}$. Equation ~\eqref{BMX} is not valid in this limit, as it implies that data is not informative for $\mathcal{M}$ anymore.
Let us interpret what Eq.~\eqref{BMX} is telling us as a Bayes factor.   One sees that $B_\mathcal{MX}$ decreases with $|\Sigma|$. It is because when the constraint $\mathcal{O}=\mathcal{O}_{ex}$ becomes more precise,  $\mathcal{M}$ is penalized, but not $\mathcal{X}$. Also, $B_\mathcal{MX}$ decreases as $|V|$ increases, because it penalizes the waste of parameter space of $\mathcal{M}$ excluded by $\mathcal{O}=\mathcal{O}_{ex}$. Finally, $B_\mathcal{MX}$ also decreases with the sensitivity $C$. $C$ measures the amount of information that $\mathcal{O}$ carries about the parameters of $\mathcal{M}$. The largest $C$ is, the more $\mathcal{O}$ contains information, and the more the constraint $\mathcal{O}=\mathcal{O}_{ex}$ is strong for $\mathcal{M}$, regardless of the experimental uncertainty.

$B_\mathcal{MX}$ is certainly useful to understand the content of the Bayesian evidence. On the other hand, it is a priori not very useful in a concrete application, as it will just tell us that the model under consideration is less good than the ideal model. Does it provide a good basis to give an absolute interpretation to $C$? The interpretation of $C$ would inevitably depend on $|\Sigma|$. It would be necessary that $|\Sigma|$ be intrinsically bounded from below, independently on the details of the experimental observations. This can actually happen, for instance in quantum theories when observables do not commute. But, as far as we know, nothing of this kind happens in a domain of physics having naturalness issues.

\subsubsection{$B_{\mathcal{MP}}$}

Consider now the Bayes factor comparing the model $\mathcal{M}$ to a  model $\mathcal{P}$ in which the observables $\mathcal{O}_{1\ldots m}$ (or any linear combination of them) are directly input parameters, such that $C=1$. It is defined by 
\be
B_{\mathcal{MP}}=\frac{p(\mathcal{O}=\mathcal{O}_{ex}|\mathcal{M})}{p(\mathcal{O}=\mathcal{O}_{ex}|\mathcal{P})}~.
\ee
 We recall that the model $\mathcal{P}$ is dubbed as ``puzzle'', since from the point of view of sensitivity, it is at the limit between predictivity and fine-tuning. In this case, the prior density for $\mathcal{O}$ will enter in the game, as $\mathcal{O}$ itself is an input parameter of $\mathcal{P}$.
 
The Bayesian evidence of $\mathcal{P}$ is
\be
p(\mathcal{O}=\mathcal{O}_{ex}|\mathcal{P})=\mathcal{L}_{max} \frac{|\Sigma|^{1/2}}{|V_{\mathcal{O}}|^{1/2}}~.
\ee
$|V_{\mathcal{O}}|^{1/2}$ is the prior volume associated to the parameter $\mathcal{O}$. 
 In this expression, one can introduce the ratio $\mathcal{O}/\mathcal{O}_{ex}$, as we did for the Bayesian evidence of $\mathcal{M}$, $p(\mathcal{O}=\mathcal{O}_{ex}|\mathcal{M})$, given in Eq.~\eqref{evidence_laplace}. By doing so, $\Sigma$ is  the same relative uncertainty $\Sigma=-\partial^2\log\mathcal{L_O} / \partial \mathcal{\log O}^i \partial \mathcal{\log O}^j$ as the one which appears in $p(\mathcal{O}=\mathcal{O}_{ex}|\mathcal{M})$.  The two $|\Sigma|^{1/2}$ thus cancel in $B_{\mathcal{MP}}$, such that
\be
B_\mathcal{MP}=\frac{ |V_{\mathcal{O}}|^{1/2}  }{ |V|^{1/2} } \int_{\mathcal{D}_{ex}}\frac{1}{C}d\sigma(\theta)~.\label{BMP}
\ee
With this choice, $C\propto\partial \log \mathcal{O} / \partial \ldots$, and the prior volume $V_\mathcal{O}$ is dimensionless. $V_\mathcal{O}$ is however not determined. To do so, the prior of $\mathcal{O}$ would need to be specified.

To go further and specify a particular  prior for $V_\mathcal{O}$, it is necessary to impose a condition, referring to some principle. Interestingly, there are two different principles, leading to two different conditions, which lead to the same result. Firstly, one can invoke the principle of indifference--introduced in Section \ref{sect:modelcomp}. Applied to a dimensionful quantity, it states that our a priori degree of belief should not depend on the unit scale. This is translated as the invariance of $p(\mathcal{O}|\mathcal{P})$ under the transformation $\mathcal{O}\rightarrow \mathcal{O}\times b$, which imposes the logarithmic prior  $p(\mathcal{O}|\mathcal{P})\propto\mathcal{O}^{-1}$.

But there is a second principle which gives the same result. In this section, we restricted our discussion to a dimensionful $\mathcal{O}$. However, there is no specification made about the actual dimension of  $\mathcal{O}$. It seems legitimate to require that the whole approach leads to the same outcome whatever the dimension of $\mathcal{O}$ is. Said differently, we require that the measure of naturalness should not depend on a redefinition of $\mathcal{O}$ changing its dimension. We will design this property as ``consistency'' of the naturalness measure. It is translated as the invariance of $B_\mathcal{MP}$ under the transformation $\mathcal{O}\rightarrow \mathcal{O}^a$. The consequence of imposing this condition is once again that $p(\mathcal{O}|\mathcal{P})$ be the logarithmic prior $p(\mathcal{O}|\mathcal{P})\propto\mathcal{O}^{-1}$.

We thus find that the logarithmic prior is independently  motivated by the principle of indifference and by the consistency of the measure. This  consistency condition is a kind of principle of indifference, applied to a Bayes factor instead of a probability. Depending on the point of view adopted, one can either claim that the consistency of the measure leads automatically to an objective prior, or that the principle of indifference leads automatically to a consistent measure.  In any case, $B_\mathcal{MP}$ is finally invariant under the transformation $\mathcal{O}\rightarrow b \times \mathcal{O}^a$. 

Provided that prior volumes of both $\mathcal{M}$ and $\mathcal{P}$ can be bounded, $B_{\mathcal{MP}}$ provides a kind of absolute scale to $C$. As expected, the ``puzzle'' model $\mathcal{P}$ plays the role of a reference in terms of sensitivity, to which $\mathcal{M}$ can be compared. Once the volumes are determined, $C$ is directly related to Jeffreys' scale. 
The interpretation of $C$ in terms of degree of belief does not depend on the definition of $\mathcal{O}$.
Indeed, any redefinition of $\mathcal{O}$ is accompanied with a change in $|V_\mathcal{O}|^{1/2}=\int d\log \mathcal{O}$, such that the interpretation of $C$ remains always the same. For instance, in the gauge hierarchy problem, it does not matter anymore to take $\mathcal{O}=m_Z$ or $\mathcal{O}=m_Z^2$, because this is compensated in the  prior volume of $m_Z$, $\int d\log m_Z\rightarrow\int d\log m^2_Z$. This consistency property resolves one of the issues raised in Section \ref{sect:sensitivity}.

\subsubsection{Relative naturalness}

Finally, let us compare two hypothesis $(\mathcal{M}_0,d_0)$ and $(\mathcal{M}_1,d_1)$. We make the assumption that the piece of data $\mathcal{O}=\mathcal{O}_{ex}$ is informative for both models. The Bayes factor 
\be
B_{01}=\frac{p(\mathcal{O}=\mathcal{O}_{ex}|\mathcal{M}_0,d_0)}{p(\mathcal{O}=\mathcal{O}_{ex}|\mathcal{M}_1,d_1)}
\ee
takes the form
\be  
B_{01}=\frac{ |V_1|^{1/2} }{ |V_0|^{1/2} } {\int_{\mathcal{D}_{0\,ex}}C_0^{-1} d\sigma(\theta)}\left(\int_{\mathcal{D}_{1\,ex}}{C_1}^{-1} d\sigma(\theta)\right)^{-1}~.
\ee
Here, one can see clearly that this naturalness measure puts in balance both the prior volumes, and the sensitivities integrated over the parameter space. 
$\mathcal{M}_0$ and $\mathcal{M}_1$ can be two different models, or the same model with different priors, or associated to different data $d_0$ and $d_1$. 
When the two models are the same, one has $C_0=C_1$, but the two domains of integration are still different.
If the two hypothesis differ only by the data $d_0\neq d_1$, $B$ gives the ``naturalness price'' between the two pieces of data. For example, $d_0$ could be the pre-LHC constraint on SUSY particle masses, and $d_1$ the constraint once LHC measurements are taken into account.

Within a same model $\mathcal{M}$, one can make the choice of punctual priors, which select two different points $\theta_0$, $\theta_1$ of $\mathcal{D}_{ex}$. In that case, the prior volumes cancel, leaving only the Bayes factor
\be  
B_{01}=\frac{C_1}{C_0}~.
\ee
The Bayes factor, then,  is simply reduced to the comparison of the sensitivities. 
This quantity shows clearly that the relative sensitivity within a model has to be interpreted on the basis of Jeffreys' scale. It is also true for the usual $c$ measure, which is a particular case of $C$. This finishes resolving all the issues raised in Section \ref{sect:sensitivity}.

\subsection{Generalization and comments}
To sum up, the Bayes factor $B_\mathcal{MP}$ provides a handle on the absolute interpretation of $C$, and sets the functional form of all quantities, once either the consistency or the indifference principle are required. The relative versions of $B$ contain various possibilities of applications,
some of them being reminiscent of previous work done in the literature. We will now discuss about generalization, absolute naturalness and some implications of this approach.

\subsubsection{The general case}

Before all, let us explain why we keep a number $m$ of observables $\mathcal{O}_{1\ldots m}$ smaller or equal to the number of parameters $n$. If one has $n$ observables, non-proportional to each other, the model is fully constrained, i.e. $\mathcal{D}_{ex}$ has dimension zero. The likelihood function gets in that case one or several maxima, with an uncertainty $\Sigma$ associated to each of them. If one adds a new constraint, the effect will be to increase the precision, i.e. reduce $\Sigma$, and possibly decrease the maximum of likelihood, if this new constraint is not in agreement with the $n$ previous. But from the point of view of the $C$ measure, this new constraint is  necessarily reduced, around the maximum, to a linear combination of the $n$ others. For that reason, the contribution of this new constraint vanishes in the determinant contained in the Jacobian factor $C$, and thus cannot influence the sensitivity. Therefore, for the purpose of the naturalness study, it is sufficient to keep $m\leq n$.


The second assumption we made in the beginning of Sect. \ref{sect:Bayes_FT} was that our observables and parameters were dimensionful.  We found that either applying the indifference principle or requiring consistency of the naturalness measure  leads to the invariance of $p(\mathcal{O})$ under $\log\mathcal{O} \rightarrow \log\mathcal{O}+b$ and of $B_\mathcal{MP}$ under $\log\mathcal{O} \rightarrow a\times\log\mathcal{O}+b$, where $a$ is a $m\times m$ matrix and $b$ a $m$-vector. 
These conditions imply the use of the logarithmic prior, such that $C= |\det(J_\mathcal{\log O} J_\mathcal{\log O}^t)|^{1/2}$, where $J_\mathcal{\log O}=\partial \log \mathcal{O}_i/\partial \log\theta_j$, and $V_\mathcal{O}=\int d^m\log\mathcal{O}_i$, $V=\int d^n\log\theta_j$.
All these properties are the consequences of considering that a transformation law, the change in unit scale, is irrelevant for our degree of belief about the problem. 
Let us now go to the general case, by considering arbitrary, possibly dimensionless, observables and parameters. All the results can be generalized, provided the existence of an irrelevant transformation. Let us assume that the transformation $G(\mathcal{O})\rightarrow G(\mathcal{O})+b$ and the transformation $H(\theta)\rightarrow H(\theta)+c$ do not modify our degree of belief. Then, the naturalness measure is invariant under  $G(\mathcal{O}) \rightarrow a\times G(\mathcal{O})+b$, the sensitivity takes the form $C= |\det(J_{G(\mathcal{O})} J_{G(\mathcal{O})}^t)|^{1/2}$, where $J_{G(\mathcal{O})}=\partial G (\mathcal{O})_i/\partial H(\theta)_j$,  and the prior volumes are $V_\mathcal{O}=\int d^m G(\mathcal{O}_i)$, $V=\int d^n H(\theta_j)$. 

What we stated above is based on the existence of a continuous irrelevant transformation. However, other kinds of conditions, possibly less obvious, can also be found. For instance, when a theory is isomorphic to itself under a duality transformation,  it is possible to find the objective priors of parameters transforming non trivially under the duality. 

Finally, it is important to recall  that results obtained in Bayesian statistics depend to some extent on the parametrization of the problem. The choice of parametrization is somehow intricate with the choice of prior for the parameters. The indifference principle (Sect. \ref{sect:modelcomp}) plays a crucial role with respect to this issue. It  allows us to minimize the amount of information contained in the priors, or, said differently, it helps to find a preferred, objective parametrization. For example, it happens that a dimensionless parameter, whose objective prior is unknown, can be seen as a ratio of two dimensionful parameters. This is for instance the case of $\tan\beta\equiv v_u/v_d$ in the MSSM. Given that the objective prior of dimensionful parameters is known, this provides the (non-trivial) objective prior of the dimensionless parameter. Or equivalently, one can choose these dimensionful parameters as input (we refer to \cite{Allanach:2007qk} for an application to the MSSM).

By construction, our Bayesian approach to naturalness inherits all of these features. However, an extra subtlety is that there is both a freedom of parameterization on the parameters $\theta_i$ and the observable $\mathcal{O}$.
Without referring to the indifference principle, there is no mean to favor a particular parametrization, and the naturalness measure is dependent on this parametrization. Once applied, the indifference principle provides  the objective priors for both the $\theta_i$ and  $\mathcal{O}$ (as discussed in the analysis of $B_{\mathcal{MP}}$). As a result, we end up with a unique naturalness measure, depending only on the transformations properties associated with the indifference principle. The general result is given above in this section.
Speaking more formally, the indifference principle defines an equivalence class among the parametrizations, and the naturalness measure turns out to be an invariant of this equivalence class.
In the usual example of dimensionful parameters, the transformation definining the equivalence class is
 $\mathcal{O}\rightarrow b \times \mathcal{O}^a$. This is nothing but the 2d set of quantities of arbitrary dimensions. As a consequence, whatever the dimension of $\mathcal{O}$, e.g. $\mathcal{O}\equiv m_Z$, $m_Z^2$, or $3\times m_Z^{100}$, the naturalness measure remains the same. 

\subsubsection{About absolute naturalness}

The `puzzle' model $\mathcal{P}$ provides a reference in terms of naturalness. It is the only sensible reference we are able to find. But how can it be defined in practice ?
Let us try to do this for the gauge-hierarchy problem.

The Bayes factor associated to this problem is \be
B_{\mathcal{MP}}=\frac{p(m_Z=m_{Z\,ex}|\mathcal{M},d)}{p(m_Z=m_{Z\,ex}|\mathcal{P},d')}\label{BF_mZ}\,.\ee
The pieces of data $d$ and $d'$ have to be identical, as our goal is not to compare different data. Which information is contained in $d$ ? By construction, in our approach, all experimental information available is splitted into two categories. There is the one which contributes to indicate what the electroweak scale is, which is called $m_Z=m_{Z\,ex}$, and the one which doesn't, which is called $d$. With only the knowledge $d$, one would know for example the strength of gravity and gauge interactions, the fermion and hadron masses, but not the electroweak boson masses neither the Fermi constant. Such a situation is of course impossible to imagine in practice, but here we are simply splitting a set of existing information, regardless of the way they were obtained.

Now, what should $\mathcal{P}$  be ? It is a model which predicts data $d$ and has $m_Z$ both as an input and an output. We can imagine that it is a kind of quantum field theory in which the weak scale does not receive any quadratic corrections, for some unknown reason. What should be the prior volume of $m_Z$ ? We know both from the indifference principle and consistency of the measure that $m_Z$ should have a logarithmic prior. The bounds of this density remain to be found. Given that $d$ contains the knowledge of gravity, $\mathcal{P}$ has a cutoff at the Planck mass, so $m_Z\leq M_{Pl}$. On the other hand, as $d$ contains the quark masses, $m_Z$ is bounded from below due to unitarity of quark scattering by weak currents (see \cite{Chanowitz:1978uj}), which implies roughly $m_Z\gtrsim 10\,\textrm{GeV}$. The prior volume $V_{m_Z}$ in the model $\mathcal{P}$ is therefore $V_{m_Z}=\log(M_{Pl}/10 \,\textrm{GeV})\approx 40.0 $. 
This completes the definition of $\mathcal{P}$. Using Laplace approximation, the Bayes factor is
\be
B_\mathcal{MP}=\frac{ |V_{m_Z}|^{1/2}  }{ |V|^{1/2} } \int_{\mathcal{D}_{ex}}\frac{1}{C_{m_Z}}d\sigma(\theta)~,
\ee
where $C_{m_Z}\propto\partial \log m_Z / \partial \ldots$ , and $|V|$ is the prior volume of $\mathcal{M}$. With this equation, for any choice of $\mathcal{M}$, $\int_{\mathcal{D}_{ex}}C_{m_Z}^{-1}d\sigma(\theta)$ is equal to Jeffrey's scale up to a known constant. Therefore we get the absolute interpretation of the sensitivity measure.

One may or not be satisfied with this approach. In any case, it illustrates that it is not so obvious to define $\mathcal{P}$ in practice. This, however, does not take away  the general results obtained by studying $B_\mathcal{MP}$.

\subsubsection{ Second order fine-tuning }

When considering a naturalness map, the following interrogation often appears. The interest of a naturalness map is to select regions of the parameter space which have a relatively low fine-tuning. But suppose that a very tiny region of the parameter space has a very small $C$, while $C$ is sensibly larger around,  in at least one direction. Selecting this tiny region and discarding the zone around would be itself an action of fine-tuning! We will design that issue as a ``second order fine-tuning''. How  is this taken into account in our framework? 

It is easy to guess that there is a relation to the choice of punctual priors, which select single points of $\mathcal{D}_{ex}$. Indeed, if $1/C$ was integrated around the tiny zone with small $C$, the particularity of that zone would disappear. Formally, the action of selecting regions with low fine-tuning corresponds to impose a prior such that $C\leq C_{min}+\Delta C$, where $\Delta C$ is a level of tolerance, and $C_{min}$ is a minimal value. 
One can construct a Bayes factor comparing two regions $\mathcal{D}_{0\,ex}$, $\mathcal{D}_{1\,ex}$ of the parameter space, containing the minima $C_{0\,min}$, $C_{1\,min}$,  respectively, and with the requirement of an upper bound on $C$. Several versions can be built, depending for example whether $C_{min}$ is considered as a common value, or if $C_{min}=C_{1,0\,min}$ respectively. These versions corresponds to different reasonings. Provided that $C^{-1}$ can be approximated over the domain considered, such Bayes factors can be computed analytically. 

For example, let us consider the Bayes factor comparing two regions $\mathcal{D}_{0\,ex}$, $\mathcal{D}_{1\,ex}$ of the parameter space, containing  minima $C_{0\,min}$, $C_{1\,min}$ which are not on the boundaries. We impose the condition $C\leq C_{min}+\Delta C$, where $C_{min}$ is a common value. It can be $\min({C_{0,\,min},C_{1,\,min}})$ or a smaller value. It does not matter, since it will not appear in the final result. These two domains are denoted as $\mathcal{D}_{0\,ex}'$, $\mathcal{D}_{1\,ex}'$. The Bayes factor reads 
\be  
B_{01}= {\int_{\mathcal{D}_{0\,ex}'}C^{-1} d\sigma(\theta)}\left(\int_{\mathcal{D}_{1\,ex}'}{C}^{-1} d\sigma(\theta)\right)^{-1}~.
\ee
When $\Delta C$ is not too large, one can take the Laplace approximation of $C$ around $C_{0\,min}$ and $C_{1\,min}$. In that limit, the integration can be done, and one obtains the Hessian of $\log C$ 
\be H=\left.\det (\nabla_i \nabla_j \log C)  \right|_{C_{min}} \,. \ee 
As by assumption, both boundaries $\partial \mathcal{D}_{i\,ex}'$ are inside the corresponding boundaries $\partial \mathcal{D}_{i\,ex}$, the Bayes factor reduces to 
\be
B_{01}=\frac{C_{1\,min}}{C_{0\,min}} \frac{H_1^{1/2} } {H_0^{1/2}}\,.\label{steep1}
\ee
We can see that a new factor $H_1^{1/2}/H_0^{1/2}$ appears in addition to $C_{1\,min}/C_{0\,min}$. This is the quantity which renders account for the second order fine-tuning. 

More generally, for Bayes factors where the minima are on the boundaries, there will be contributions of the form 
\be
B_{01} \propto  \prod_i  \left| \frac{\partial C}{\partial \theta_i}  \right|_{C_{1\,min}}  \left| \frac{\partial \theta_i}{\partial C }  \right|_{C_{0\,min}} \label{steep2}
\ee
coming in. Terms in Eqs~\eqref{steep1},\eqref{steep2} provide a comparison of the steepness of $C$ around the two minima. This is properly quantified and interpreted in terms of naturalness with the Bayesian approach.


\subsubsection{The top Yukawa in the gauge hierarchy problem}

A recurring question about the gauge hierarchy problem is whether or not the top quark Yukawa coupling $y_t$ should be considered as an input parameter, such that the derivative $\partial m_Z / \partial y_t$ appear in the $C_{m_Z}$ measure. On one hand, one can think of $y_t$ as a simple constant, and not a parameter. In that case, it should not appear in $C_{m_Z}$. On the other hand, one can think of it as an input parameter, fixed by the experiment. In that case, $y_t$ must appear in the $C_{m_Z}$ measure. So what is the right point of view? Surprisingly, it is the first proposition which makes sense. To understand this, we have to examine more carefully the second proposition. 

Indeed, the choice of considering $y_t$ as an input parameter or a constant is just a matter of viewpoint, and should not modify the information content of our study. This implies that if $y_t$ is taken as an input parameter, one has to add to the set of experimental constraints the top quark mass measurement, $m_t=m_{t\,ex}$. But the observables $m_Z$ and $m_t$ are not independent in the model. Therefore, to study naturalness of the gauge-hierarchy problem, they need to be  simultaneously taken into account. It is thus the combined sensitivity $C_{m_Z,\,m_t}$  which has to be used when $y_t$ is seen as an input parameter. 

%
%
%
%
%
%
%
%
%

At this point, it is instructive to wonder what is the common fine-tuning associated to a generic observable $\mathcal{O}$ and an observable $\Theta$ which is directly an input parameter. The set of the input parameters is denoted as $p_i=(\theta_j, \Theta)$. We assume a logarithmic prior for all quantities for concreteness. 
 If $\mathcal{O}$ and $\Theta$ are independent in the model, the common sensitivity 
\be C_{\mathcal{O},\Theta}= \left\|\frac{\partial \log \mathcal{O}}{\partial \log p_i }  \wedge \frac{\partial \Theta}{\partial \log p_i }\right\| \ee 
factorizes and reduces to \be C_{\mathcal{O}}=\left\| \frac{ \partial \log \mathcal{O}}{\partial \log \theta_i }   \right\| \,,\ee given that $C_\Theta=1$. But what happens when the two observables are correlated? It turns out that the answer is the same. Whatever the ``puzzle'' observable $\Theta$ is, the sensitivity reduces always to $C_{\mathcal{O},\Theta}=C_{\mathcal{O}}$. We emphasize that, although all priors are chosen to be logarithmic there, these kinds of results hold whatever the priors are.

Let us come back to the top Yukawa and the $C_{m_Z,\,m_t}$ measure. The previous remark does not apply directly, because the observable is not $y_t$, but rather the top mass $m_t=y_t\times v$. Thus $y_t$ does not play the same role as $\Theta$. However, the outcome will in fact  be the same. 
We denote the set of input parameters as $p_i=(\theta_j, y_t)$. The objective prior of $m_t$ is logarithmic, and it implies that the prior of $y_t$ is also logarithmic. We also assume for simplicity that the priors of the $\theta_j$ are logarithmic. The sensitivity is then \be \left\|C_{m_Z,\,m_t}=\frac{\partial \log m_Z}{\partial \log p_i }  \wedge \frac{\partial \log (v y_t)}{\partial \log p_i } \right\|\,.  \ee
As $m_Z$ is directly related to $v$, the gradients $\partial \log m_Z/\partial \log p_i $ and $\partial \log v/\partial \log p_i $ are colinear. The $v$ contribution therefore vanishes in the sensitivity measure. The remaining part contains $y_t$, which  plays the same role as $\Theta$  in the previous paragraph. As a  consequence the sensitivity reduces to  $C_{m_Z}$, without the parameter $y_t$, \be C_{m_Z,\,m_t}=\left\| \frac{\partial \log m_Z}{\partial \log\theta_i } \right\|\,.\ee 
This result holds whatever the priors are. Thus, to reply to the initial question, the second proposition gives in fact the same result as the first proposition, after a careful examination: $y_t$ should not appear in $C_{m_Z}$.

\subsubsection{Consequences of LHC searches}

The existence of a scalar resonance whose properties are roughly compatible with the one of a Higgs boson has been established beyond reasonable doubt at the LHC \cite{cms_higgs, atlas_higgs}. The mass of this new state is a a stringent constraint on many models of new physics. Some of them are almost excluded, baring some very specific choices of parameters. As a result, this constitutes a new naturalness problem associated to the Higgs mass constraint. It would be therefore particularly appropriate to study the fine-tuning related to the Higgs, either inside the parameter space of a model, or comparing two different models. The naturalness measure to use for such study is \be B_{01}=\frac{p(m_h=m_{h\,ex},m_Z=m_{Z\,ex})|\mathcal{M}_0)}{p(m_h=m_{h\,ex},m_Z=m_{Z\,ex})|\mathcal{M}_1)}\,.
 \ee\,
We emphasize once again that the two observables $m_h$ and $m_Z$, both independently responsible of some amount of fine-tuning, should not be treated separately, because their predictions are correlated in the models.

On the other hand, searches  at the LHC and other experiments do not have, up to now, shown conclusive evidence of existence of Beyond Standard Model phyics.
As the idea of new physics (NP) at the multiTeV scale is in part motivated by the gauge-hierarchy problem, one can wonder to which extent NP models are more natural than the Standard Model, given the increasing exclusion limits.
Let us answer to this question in a very simplified way. For concreteness, we assume the SM to be valid up to the Planck scale. As the origin of the gauge-hierarchy problem is an issue of cancellations between square mass parameters, we will consider a one-parameter ``model'' embedding this property. That is,  we just define the EW scale as given by   $m_Z^2=M_{Pl}^2(1-\delta)$. 
We also consider a BSM model suppressing the quadratic corrections to the EW scale at a scale $\tilde{M}<M_{Pl}$. The EW scale is thus given by $m_Z^2=\tilde{M}^2(1-\delta)$ in this model. Picking similar priors for the $\delta$ in each hypothesis, the naturalness measure turns out to be
\be
B_{NP,SM}\approx \frac{M_{Pl}^2}{\tilde{M}^2}\,.
\ee
We can see that, unless $\tilde{M}$ is close from $M_{Pl}$, this ratio indicates an extremely strong fine-tuning of the SM, far beyond the $150$ typical value indicated in Jeffrey's scale. If for instance $\tilde{M}\approx 100\,\textrm{TeV}$, one gets $B_{NP,SM}\approx 10^{26}$.
We emphasize that, for the sake of comparing the SM to a NP model improving substantially the gauge-hierarchy problem, there is no need to set up a more evolved analysis. This estimation  embeds the large leading  contribution, which flushes away any other subleading effects.

\section{Gauge hierarchy problem and neutralino dark matter in the cMSSM }\label{sect:SUSY}

In this section, we apply our results to a concrete problem. We choose to study the naturalness of a classic supersymmetric model, the constrained MSSM (cMSSM), taking into account both the gauge hierarchy problem and the fine-tuning of neutralino dark matter. 

Supersymmetry solves the gauge hierarchy problem by embedding the Standard Model fields into supermultiplets, which do not generate quadratic corrections to the Higgs mass. The simplest realistic model this one can build is called the Minimal Supersymmetric Standard Model (MSSM). But the superparticles which accompany the SM particles in the supermultiplets are experimentally constrained to be heavier than their standard partner. This implies that supersymmetry has to be broken. However, with broken SUSY, the gauge-hierarchy problem is not completely solved.  Instead, it remains in the form of a certain amount  of special cancellations between the SUSY parameters, of typical scale $M_{SUSY}$,  necessary  to reproduce the Z boson mass. $M_{SUSY}$ is constrained both through direct and indirect observations, and the LHC experiments  are currently improving these direct limits (see e.g.  summary plots of Atlas \cite{ATLAS} and CMS \cite{CMS}). Roughly speaking, $M_{SUSY}$ is at least $O(\textrm{TeV})$, one order of magnitude above the Z mass $m_Z\approx 91~  \textrm{GeV}$. 

One of the simplest and widely studied version of the MSSM with broken SUSY is the constrained MSSM (cMSSM). The parameters of that model are a common gaugino mass $m_{1/2}$,  a common scalar mass $m^2_0$, a common scalar trilinear coupling  $A_0\equiv a_{ij}/y_{ij}$, the ratio of the two Higgs vevs $\tan\beta=\left\langle H_u \right\rangle/\left\langle H_d \right\rangle$, and the sign  of the SUSY Higgs mass term, $\textrm{sign}(\mu)$ (see e.g. \cite{Martin:1997ns} for an introduction to SUSY models). However, this is a setup which already takes into account $m_Z=m_{Z\,ex}$. As we are interested in the fine-tuning induced by $m_Z=m_{Z\,ex}$, this constraint must not be incorporated in the model in the first place. Therefore a new input parameter has to be introduced. It is in fact interesting to trade $\tan\beta$ for the dimensionful parameters $\mu$ and $B_\mu$ of the Higgs sector. Indeed, whereas it is not obvious to find an objective density for $\tan\beta$, the objective densities of $\mu$ and $B_\mu$ are clearly logarithmic. In practice, it is the former parametrization of the model which is used. In that case, the objective prior of $\tan\beta$  has to be inferred from the priors of $\mu$ and $B_\mu$. This remark was made in the paper \cite{Allanach:2007qk}, where the resulting density is called ``REWSB prior''.

Also, the MSSM has another celebrated feature. Its mass spectrum contains the neutralino, a fermion charged only under the weak force, and which is a mixture of neutral Higgsinos and gauginos. If the lightest neutralino $\tilde{\chi}_1^0$ is the lightest particle of the SUSY spectrum, and if a remnant of the $U(1)_R$ symmetry of the SUSY algebra is still present, it cannot decay directly into SM particles and is therefore stable. Such a particle is a good dark matter candidate. Under the assumptions that the Cosmological Standard Model is valid in the early universe, and that the neutralinos were at the thermal equilibrium for some period, today's neutralino density can be precisely predicted using the Boltzmann equation. This density is the relic remaining after thermal freeze-out, when the neutralino annihilation rate vanishes due to the expansion of the universe.
The relic density predicted strongly depends on the masses and compositions of all particles of the spectrum.

\begin{figure}
\centering
\includegraphics[width=6.4cm]{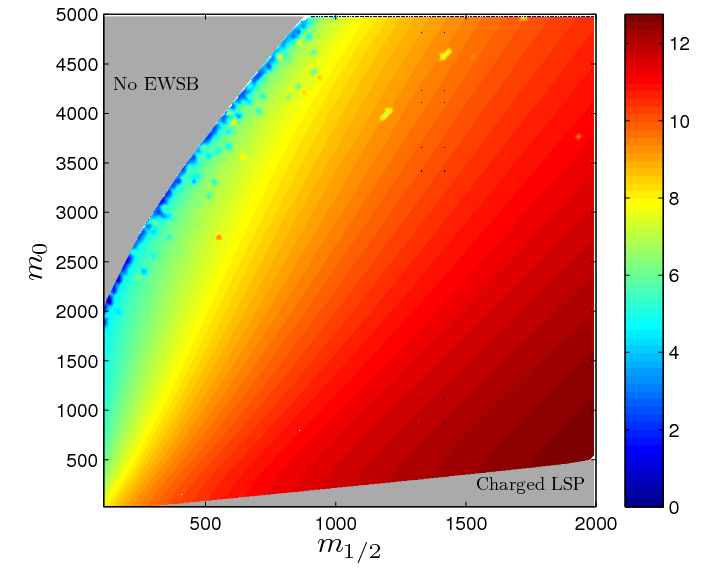}
\includegraphics[width=6.4cm]{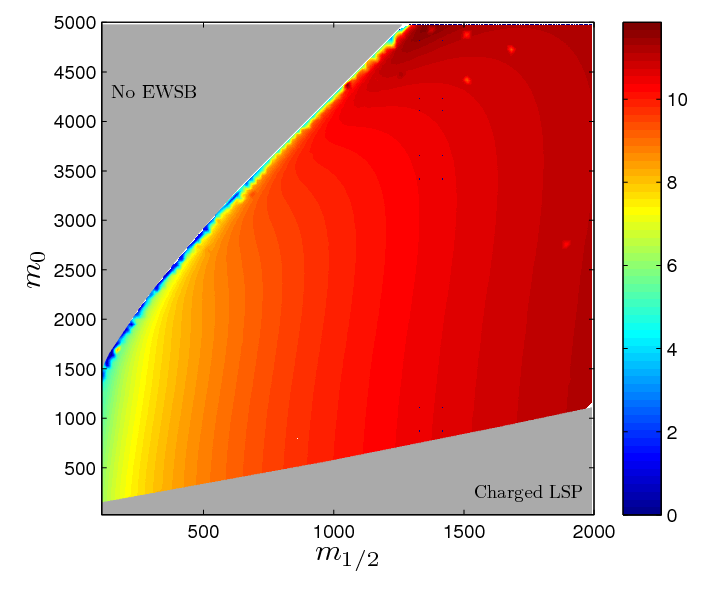}
\centering
\includegraphics[width=6.2cm]{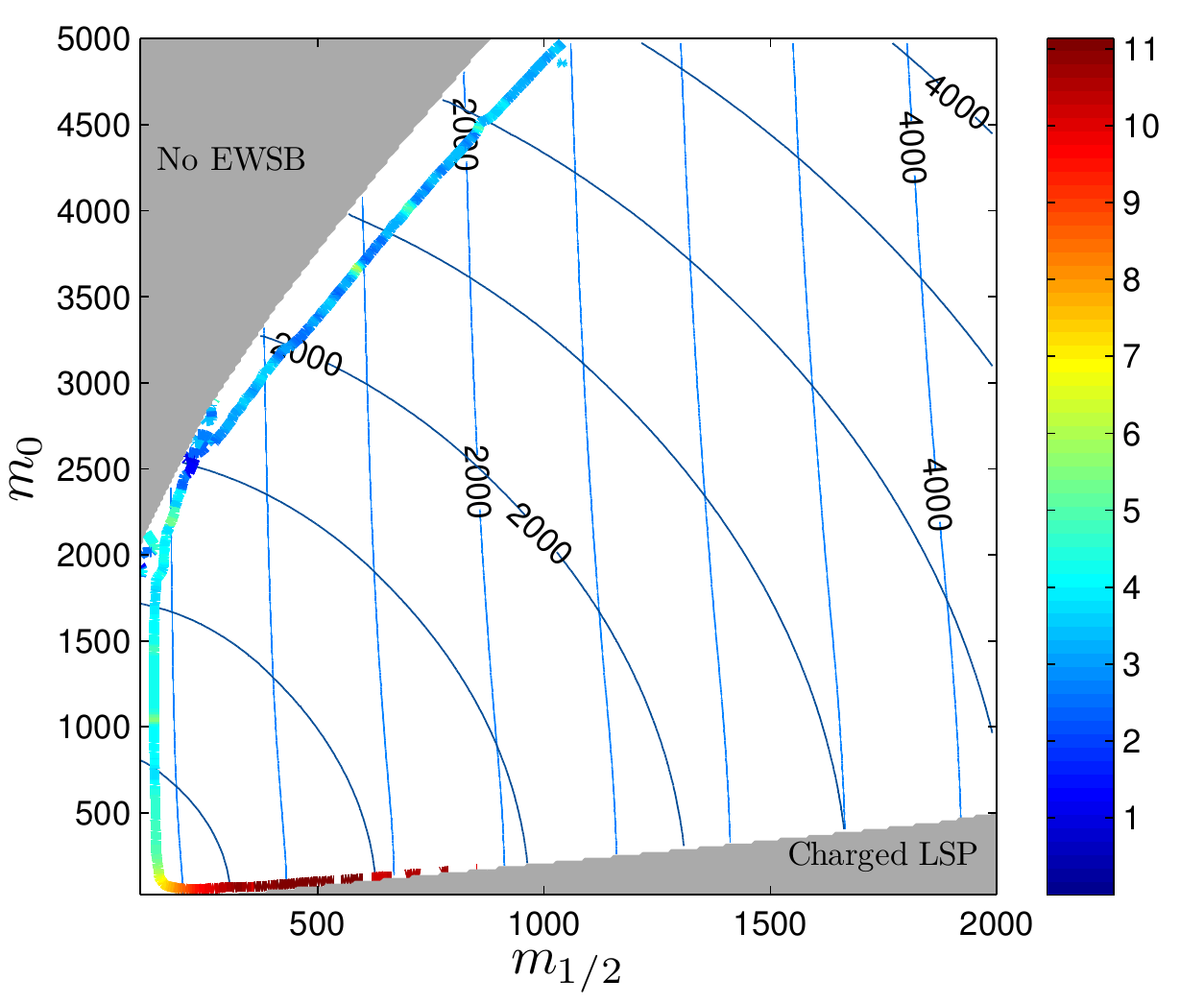}
\includegraphics[width=6.5cm]{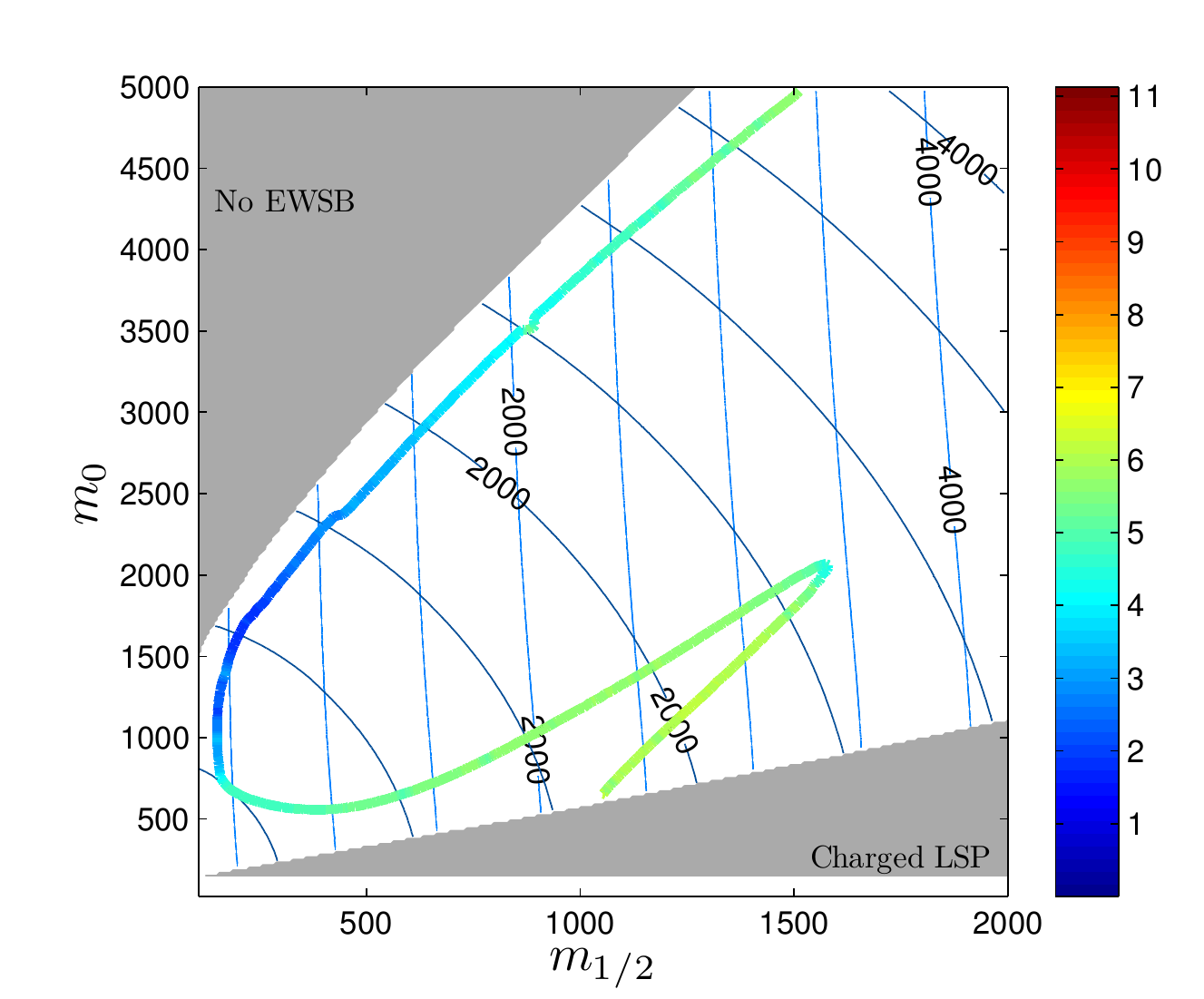}
\centering
\includegraphics[width=6.2cm]{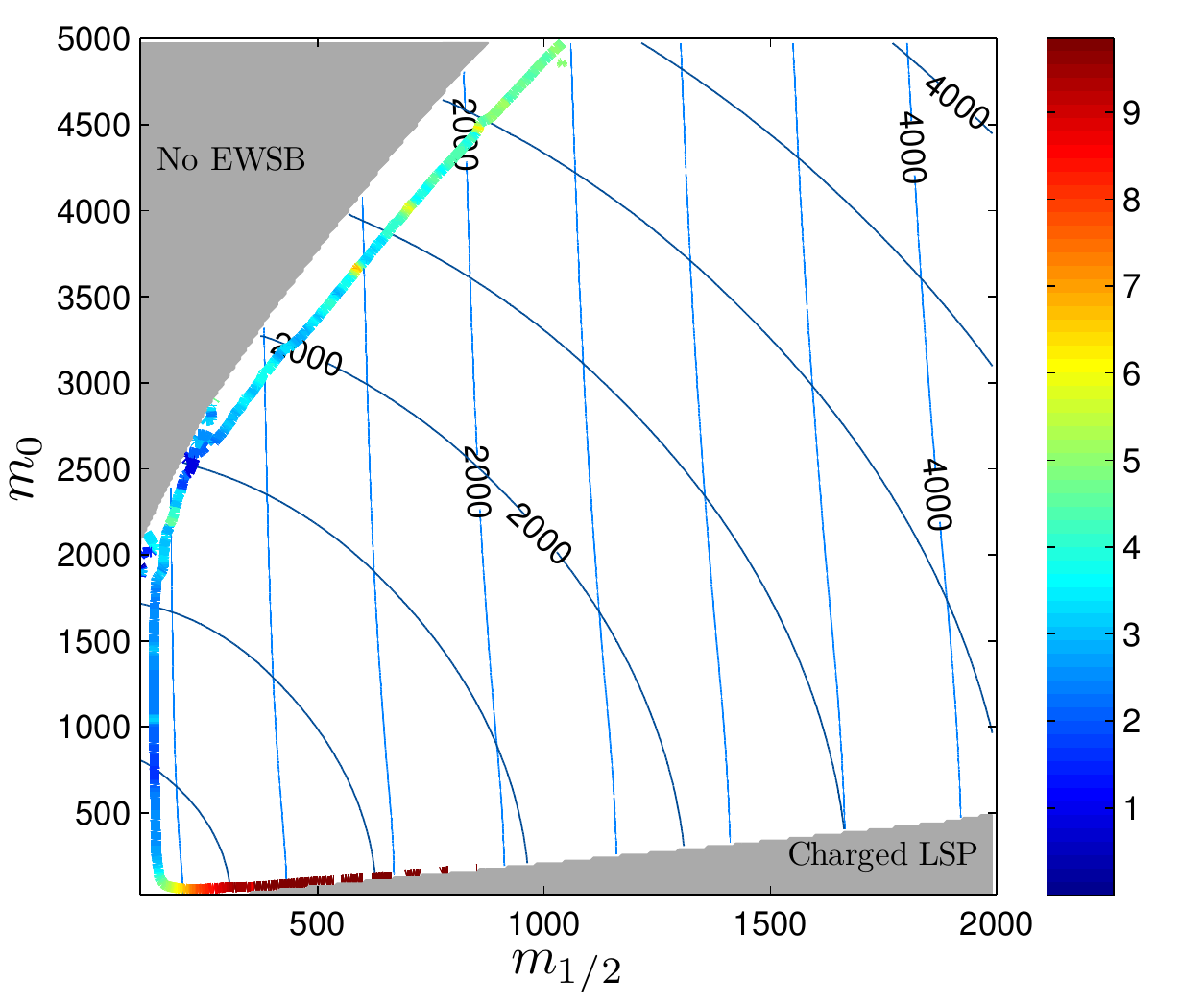}
\includegraphics[width=6.5cm]{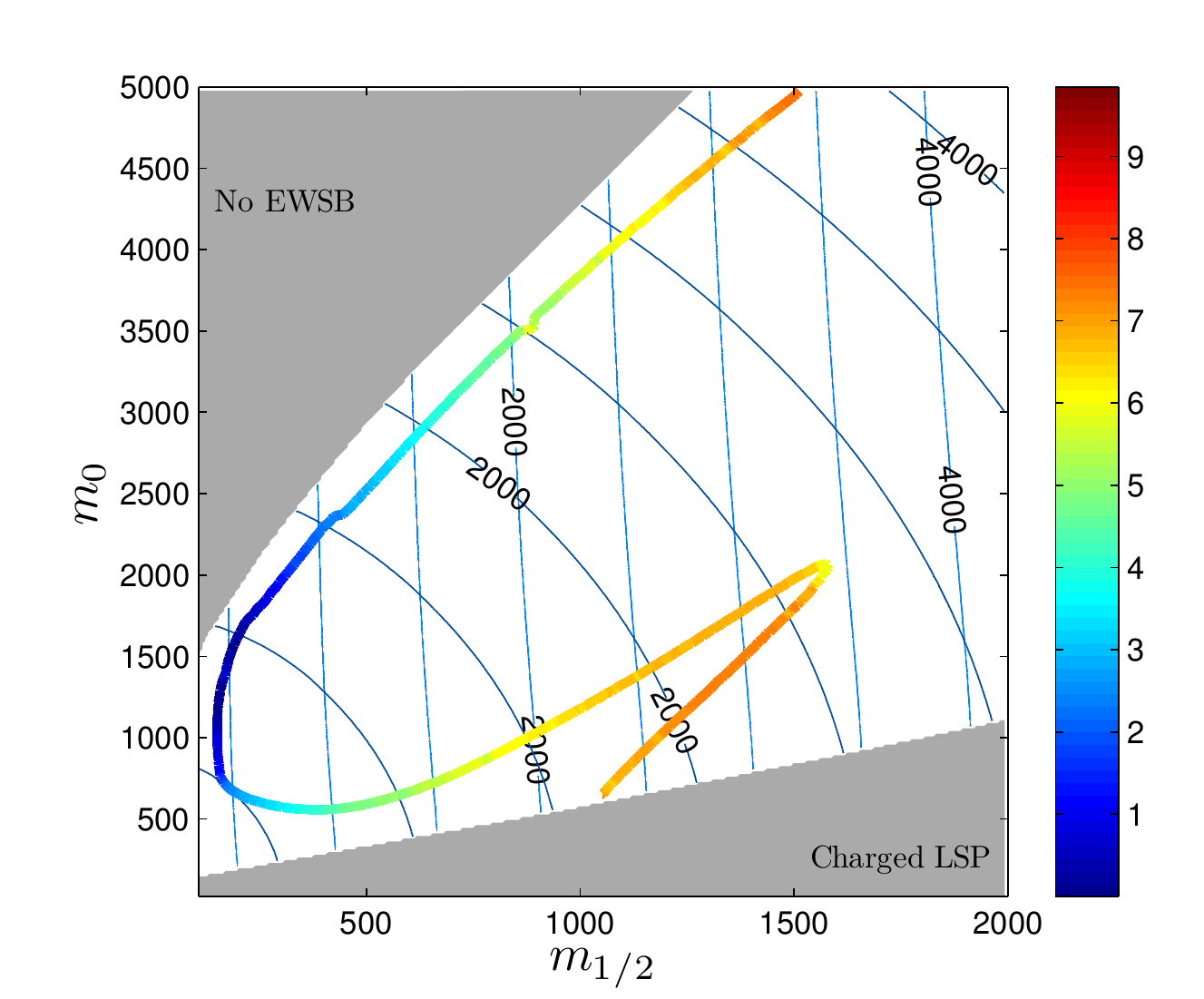}

\centering
\caption{
Quantified naturalness in the cMSSM. All of these plots are for $A_0=0$, $\textrm{sign}(\mu)=1$ and $m_{t}=172.4\,\textrm{GeV}$,  with $\tan\beta=10$ and $50$ for  left and right panels. $m_{1/2}$ and  $m_0$ are given in GeV units.
\newline
 \textit{Top row:} Maps of the logarithm of the electroweak fine-tuning measure $C_{m_Z}$, normalized to the point of minimal fine-tuning.
\textit{Center row:} Maps of the logarithm of the dark matter fine-tuning measure $C_{\Omega}$. Measure on both plots has the same normalization.
\textit{Bottom row:} Maps of the logarithm of the combined electroweak and dark matter fine-tuning measure $C_{m_Z,\,\Omega}$. Measure on both plots has the same normalization.
\newline
Blue and dark blue isolines show the mass of the gluino and the lightest squark with steps of 500 GeV.
Following Jeffreys' scale, the relative degree of belief between two points $1$ and $2$ is given by $|\log (C_2/C_1)|$, such that threshold values $1$, $2.5$ and $5$ correspond to weak, moderate and strong evidence for point $1$, respectively.
  \label{fig:plots}}
\end{figure}

The latest release of the dark matter relic density measured by WMAP7 is $\Omega h^2_{ex}= 0.1126 \pm 0.0036$ \cite{Komatsu:2010fb}. One can know  where this constraint is satisfied in the plan $(m_{1/2},m_0)$ by looking at the lines in plots of Fig. \ref{fig:plots}. This figure will be described in details below. Experimental uncertainty, by construction, does not appear in the plots. Lines are set wide only to ease the reading of the color code. 
Typically, in the cMSSM, the dark matter relic density predicted is a bit large compared to the observation.  To reproduce this experimental constraint, it is necessary that one or several processes of neutralino annihilation be particularly efficient\cite{Baltz:2006fm}. There are at least four such processes in the cMSSM.  

Firstly, the neutralinos can annihilate through the exchange of a scalar. But this mechanism is only efficient for light sparticles, which are more and more excluded by the LHC. Secondly, the annihilation through the exchange of a Higgsino or $SU(2)$ gaugino is efficient if the mixing with those states is large enough. This happens near the ``No EWSB'' zone. Thirdly, a coannihilation with a slepton may dominate if it is close from the neutralino mass, and if both particles are not too heavy. This happens near the ``Charged LSP'' zone. Finally, the exchange of a CP-odd higgs $A^0$  is enhanced near the resonance pole, when $m_{A^0}\approx 2m_{\tilde{\chi}_1^0}$. This happens at large $\tan\beta$ and is dubbed ``$A$-pole funnel''.

But relying on the efficiency of such processes to obtain the correct value for $\Omega h^2$ requires a rather precise adjustment of parameters. It is therefore an act of fine-tuning.  So if one wants to explain dark matter by the neutralino, one ends up with two naturalness problems, one induced by the piece of information $m_Z=m_{Z\,ex}$ and the other due to $\Omega=\Omega_{ex}$. To study fine-tuning in the cMSSM, it is therefore the common sensitivity, $C_{m_Z,\, \Omega h^2}$ which must be used. From the sensitivity point of view, one has to consider the set of fundamental parameters $p_i=(m_{1/2}, m_0^2, A_0, \mu, B_\mu)$. All of those parameters are defined at the GUT scale. They all have a logarithmic prior as objective density. On the other hand, although $\Omega=\rho_{CDM}/\rho_c$ is a density rescaled to be made dimensionless, $\rho_{CDM}$ is dimensionful, so it necessitates a logarithmic prior as well. The common sensitivity measure is therefore \be
C_{m_Z,\, \Omega}=  \left\| \frac{\partial\log m_Z}{\partial\log p_i}   \wedge  \frac{\partial\log \Omega h^2}{\partial\log p_i}  \right\|\,.\label{C_both}
\ee 
We assume that the experimental uncertainties are sufficiently small, such that Eq.~\eqref{C_both}  holds for all the parameter space. This sensitivity will be denoted as $C$ from now on.

In the MSSM, the top quark mass is given by $m_{t}=y_t v \sin\beta$. Thus rigorously, $y_t$ should not be taken as a constant, since what we explained in  Subsection 4.4 about the top Yukawa does not hold here due to the presence of $\sin\beta$. To stay exact, it would be necessary to consider the sensitivity associated to the three observables, $C_{m_Z,\,\Omega ,\,m_{t}}$. However, the correction induced from adding the observable $m_t$ is small, because in the $\sin\beta$ contribution the derivative $\partial \log \sin\beta / \partial \log y_t$ is dominant over the other derivatives. Therefore we choose to work only with the observables $m_Z,\,\Omega h^2$, and keep $y_t$ as a constant.

We  evaluated the dark matter relic density, the sensitivity $C$ and the SUSY spectrum over slices of the parameter space of the cMSSM. Our analysis was realized using a modified version of the spectrum calculator \verb!SoftSUSY! \cite{Allanach:2001kg} interfaced with \verb!MicrOMEGAs2.4! \cite{Belanger:2004yn} to compute the dark matter relic density. In spectrum calculators, these are not $\mu$ and $B_\mu$ which are input parameters, but $m_Z$ and $\tan\beta$. This is already taken into account in \verb!SoftSUSY! to compute the $m_Z$ derivatives, but it has to be carefully considered when implementing the $\Omega h^2$ derivatives. 
The results obtained are presented in Fig. \ref{fig:plots}. 

Figure \ref{fig:plots} shows slices of the parameter space with $A_0=0$, $\textrm{sign}(\mu)=1$ and $m_{t}=172.4\,\textrm{GeV}$,  for  $\tan\beta=10$ and $50$. 
The low mass region of the parameter space is increasingly excluded by the LHC bounds on sparticle masses (see \cite{ATLAS,CMS}). Instead of showing a particular limit, we prefer to plot the gluino and lightest squark masses, and leave the choice to the reader to apply his preferred bound. 
All the plots display logarithm of the sensitivity. Following our results of Section \ref{sect:Bayes_FT}, the difference between two points $1$ and $2$ is given by $|\log C_2-\log C_1|=\Delta \log C$, which has to be interpreted on the basis of Jeffreys' scale.
 That is, $\Delta \log C=1,\,2.5,\,5$ correspond to weak, moderate and strong evidence in favour of point $1$, respectively. The statements about naturalness that we will make when discussing the plots are based on this scale.

Plots in the upper line show maps of electroweak fine-tuning. 
The fact that $\log C_{m_Z}$  drops down near the ``No EWSB'' zone is due to a feature of the MSSM renormalization group equations, known as the mechanism of ``focus point'' \cite{Feng:2000gh}. In short, the low scale value of the Higgs soft mass $m^2_{H_u}$ becomes generically small in this region, such that the cancellations required to reproduce the Z mass are less important. This feature also implies a large higgsino fraction for the neutralino, so that the experimental value of $\Omega h^2$ can be reproduced. In this zone of the parameter space, the predictions of $m_Z$ and $\Omega h^2$ are therefore particularly correlated by the model.

Plots in the center line show the dark matter fine-tuning. In the $\tan\beta=10$ slice, one can see that the coannihilation region has a very strong fine-tuning compared to the focus point region. Formally, the coannihilation region continues all along the `` charged LSP zone '' with an increasing fine-tuning, but points are so fine-tuned that the numerical analysis does not render them. In the $\tan\beta=50$ slice, one can see that the $A$-pole funnel and the focus point region have sensibly the same naturalness. Relative to the $\tan\beta=10$ focus point, these regions have a weak to moderate fine-tuning. On the other hand, they are strongly more natural than the $\tan\beta=10$ coannihilation region. At $\tan\beta=50$, some very fine-tuned coannihilations can also occur on the border, but are not shown on the plot. Dark matter fine-tuning has been previously investigated in the literature, see e.g. \cite{Cassel:2010px, Ellis:2007by }, with slightly different definitions for $C_\Omega$.

Finally, plots shown in the lower line are for the combined electroweak and dark matter fine-tuning  measure $\log C_{m_Z,\,\Omega}$. Compared to the dark matter fine-tuning alone, here the $C$ measure increases with $M_{1/2}$ due to the gauge hierarchy problem.
In the $\tan\beta=10$ slice, the coannihilation region is still strongly fine-tuned with respect to the focus point region.
In the $\tan\beta=50$ slice, the fine-tuning of the focus point region increases by $\Delta \log C\approx5$ between $m_{1/2}=500\,\textrm{GeV}$ and $1500\,\textrm{GeV}$. At low $m_{1/2}$, this region is the most favored. At high $m_{1/2}$, it is moderately fine-tuned compared to the focus point region at $\tan\beta=10$. The $A$-pole funnel and the focus point region at high $M_{1/2}$ are only moderately preferred to the $\tan\beta=10$ coannihilation region. 

\section{Conclusion}\label{conclusion}

The degree of naturalness is often intuitively defined as a sensitivity, although this approach suffers from several conceptual flaws. We propose a different definition to formalize naturalness, working in the framework provided by Bayesian statistics. This approach is self-consistent, and interestingly, turns out to embed the usual sensitivity definition in a generalized form. 

So our approach is not an alternative. It appears that the sensitivity is actually a piece, intuitively guessed, of a larger setting. It is not consistent when taken alone, but  the flaws find an explanation once the embedding in the Bayesian framework is done. Somehow, the essential missing piece was the notion of prior volume, which is also intuitive on its own.  In this paper, we work out the consistent framework bringing together these notions. 
 
The naturalness measure which appears in this framework is a Bayes factor. The link between  the naturalness measure and our degree of belief, which was missing so far,  is therefore automatically provided by Jeffreys' scale. The generalized sensitivity which emerges  takes into account  the fine-tuning of an arbitrary number of correlated observables. We discussed in details the two observable case.

By studying the Bayes factor involving a `puzzle' model, we found that either the principle of indifference or consistency of the measure are setting the functional form of all quantities. For the sensitivity, it entails that these are the objective prior repartition functions, both for parameters and observables, which appear in the derivatives. As the `puzzle' model is a reference in terms of sensitivity, this Bayes factor gives a handle on the absolute interpretation of $C$.

The Bayesian approach resolves without ambiguity the question of whether or not  the top Yukawa should enter in the gauge-hierarchy measure $C_{m_Z}$. Also, it accounts for the ``second order fine-tuning'', which is induced when it is necessary to adjust precisely a parameter to select a zone with small fine-tuning in the parameter space. Consequences of recent LHC searches are also discussed.

We present a simple illustration of our results by examining the naturalness of a supersymmetric model, the cMSSM. The sensitivity formulas associated to the electroweak scale and dark matter relic density, taken separately or together, are well-defined, and differ from some work in the literature.
By using Jeffreys' scale, we make statements about naturalness of the different dark matter annihilation regions. Roughly speaking,  the focus point region is the winner of the naturalness comparisons, while the coannihilation region comes last with a strong evidence gap.

\section*{Acknowledgments} 

I am grateful to Sabine Kraml for her support and for collaboration in the early stages of this work. 
I would like to thank Dumitru Ghilencea, Dmitri Melnikov,  B\'eranger Dumont, Florian Lyonnet,  for useful discussions and suggestions, and particularly Suchita Kulkarni for the numerous comments about the draft. 
   
   \clearpage

\noindent{\Large\bf Appendix}

\appendix

\section{Naturalness problems in particle physics and cosmology\label{naturalness_issues}}

In this Appendix, we recall some of the main naturalness problems. These are the most commonly discussed, but the list is not intended to be exhaustive.

\begin{itemize}

\item
Gauge hierarchy problem\cite{Martin:1997ns}: The electroweak scale, often represented by the Z boson mass $m_Z$, is $O(100~\textrm{GeV})$. On the other hand, the Planck scale $M_{Pl}=\sqrt{\hbar c/8\pi G_N}=2.4\times10^{18}~\textrm{GeV}$, sets the scale at which the theory of quantum gravity appears. Why is the electroweak scale so small compared to $M_{Pl}$, while it should receive $O(M^2_{Pl})$ quantum contributions? 

\item
Strong CP puzzle\cite{Kim:2008hd}: From neutron electric dipole measurement, one deduces that the $\theta$ angle, contributing to the QCD lagrangian $\mathcal{L}_{QCD}\supset -1/4g^2 G^{\mu\nu}G_{\mu\nu}+\theta/16\pi^2 G^{\mu\nu}\tilde{G}_{\mu\nu}$, is very small, $\theta<10^{-12}$, while it could take values in $[-\pi,\pi]$. Then, why is it so close to zero?  

\item
Flavour puzzle\cite{PDG}: Ratios of successive SM fermion mass eigenvalues, as well as CKM angles, are all roughly of the same order. Why do they follow this particular structure?

\item
Cosmological constant problem\cite{Bousso:2007gp}: The cosmological constant $\Lambda$, which appears in Einstein's equations $\mathcal{R}_{\mu\nu}-\frac{1}{2}g_{\mu\nu}\mathcal{R}=8\pi G_NT_{\mu\nu}+\Lambda g_{\mu\nu}$, is estimated to be $O(10^{-47})~\textrm{GeV}^4$ by fitting the Standard Cosmological Model to CMB, large scale structures, supernovae data. Within quantum field theories, it should receive $O(M^4_{Pl})$ contributions, (or $O(M^4_{SUSY})$ if  there is SUSY). Then why is it so small?

\item
Flatness problem\cite{Baumann:2009ds}: In the Standard Cosmological Model, the curvature of universe is given by $1/R^2=H^2(\rho/\rho_c-1)$, where $H$ is the Hubble constant, $\rho$ is the total energy density contained in the universe, and $\rho_c=3H^2/8\pi G_N$. $\rho/\rho_c-1$ is estimated  to be less than $0.01$, and $O(10^{-61})$ at the Planck era. Why the universe had such a small curvature?

\item
Cosmic coincidence\cite{ArkaniHamed:2000tc}: Why are the densities of matter and vaccum energy of same order of magnitude, i.e. $\rho_M\sim\rho_\Lambda$? And why now?

\end{itemize}

\end{document}